\documentclass[twocolumn,floatfix]{aastex63}
\usepackage{algorithm}
\usepackage{algpseudocode}
\usepackage{verbatim}
\usepackage{hyperref}
\usepackage{fancyvrb}

\usepackage{xspace} % so we don't have to type {} after empty commands
\xspaceaddexceptions{]\}}
\begin{document}

\title{On the Significance of Rare Objects at High Redshift: The Impact of Cosmic Variance}
\correspondingauthor{Christian Kragh Jespersen}
\email{ckragh@princeton.edu}

\author[0000-0002-8896-6496]{Christian Kragh Jespersen}
\affiliation{Department of Astrophysical Sciences, Princeton University, Princeton, NJ 08544, USA}

\author[0000-0003-3780-6801]{Charles L. Steinhardt}
\affiliation{Cosmic Dawn Center (DAWN)}
\affiliation{Niels Bohr Institute, University of Copenhagen, Lyngbyvej 2, DK-2100 Copenhagen \O}

\author[0000-0002-6748-6821]{Rachel S. Somerville}
\affiliation{Center for Computational Astrophysics, Flatiron Institute, 162 5th Avenue, New York, NY 10010, USA}

\author[0000-0001-7964-5933]{Christopher C. Lovell}
\affiliation{Institute of Cosmology and Gravitation, University of Portsmouth, Burnaby Road, Portsmouth, PO1 3FX, UK}
\affiliation{Astronomy Centre, University of Sussex, Falmer, Brighton BN1 9QH, UK}

\begin{abstract}
The discovery of extremely luminous galaxies at ultra-high redshifts ($z\gtrsim 8$) has challenged galaxy formation models. Most analyses of this tension have not accounted for the variance due to field-to-field clustering, which causes the number counts of galaxies to vary greatly in excess of Poisson noise. This super-Poissonian variance is often referred to as cosmic variance. Since cosmic variance increases rapidly as a function of mass, redshift, and smaller observing areas, the most massive objects in deep \textit{JWST} surveys are severely impacted by cosmic variance. We construct a simple model, including cosmic variance, to predict the distribution of the mass of the most massive galaxy for different surveys, which increases the tension with observations. The distributions differ significantly from previous predictions using the Extreme Value Statistics formalism, changing the position and shape of the distributions. We test our model using the \texttt{UniverseMachine} simulations, where the predicted effects of cosmic variance are clearly identifiable. We find that the high skew in the distributions of galaxy counts for typical deep surveys imply a high statistical variance on the cosmic variance itself. This impacts the calibration of the cosmic variance, as well as the expected mass of the most massive galaxy. We also find that the impact of cosmic variance dominates the impact of any realistic scatter in the stellar-to-halo-mass relation at $z\gtrsim 12$. It is therefore crucial to accurately account for the impact of cosmic variance in any analysis of tension between early extreme galaxies and galaxy formation models.
\end{abstract}


\section{Introduction}

\label{sec:intro}

In the $\Lambda$CDM paradigm, structure forms hierarchically, with perturbations in the matter distribution collapsing and merging to form progressively larger and more massive structures. Current galaxy formation models build on top of this, modelling baryons as falling into the potential wells formed by the collapsed dark matter halos, cooling and turning into stars over time \citep{SomervilleDave_2015}. Recent observations from the James Webb Space Telescope \citep[\textit{JWST},][]{Gardner2006_JWST} have detected a population of luminous galaxies at extremely high redshifts $z\gtrsim 10$, with both number densities and masses significantly higher than those predicted by pre-launch theoretical models \citep[e.g.][]{Harikane23, Leung2023, Finkelstein2024}. Although there were early claims that these galaxies were in fundamental tension with the $\Lambda$CDM paradigm \citep{Boylan-Kolchin23_early_galaxies, Labbe23_red_early, Parashari2023_highz_primordial_power_spectrum}, it is now generally accepted that the observed galaxies can be accounted for within the standard $\Lambda$CDM paradigm if the effective conversion of baryons into stars is higher than that in the nearby Universe \citep{Dekel2023_bursts}. Other possible difficulties in interpreting such observations are the very large uncertainties on stellar mass estimates, as well as additional uncertainties on the contribution from accreting black holes to the observed rest-UV luminosity, and possible evolution in the stellar initial mass function \citep{Steinhardt2023_IMF_template}. However, the unexpectedly high masses of the most massive galaxies are still a potential challenge for galaxy formation models. 

The analyses of the most massive galaxies to date have mostly focused either on the \textit{expected} (average) number of galaxies of a certain mass at a given redshift \citep{Boylan-Kolchin23_early_galaxies}, or on the Extreme Value Statistics (EVS) technique \citep{Lovell2023_EVS}. However, both approaches neglect field-to-field clustering due to large-scale structure. 
Field-to-field differences due to clustering are highly significant for the smaller pencil-beam \textit{JWST}-surveys at higher redshifts \citep{Moster2011}, especially for the highest-mass galaxies, which cluster the strongest, a fact that has long been noted in high-z fields \citep{Adelberger1998_cosmic_variance_highz, Robertson2010_measure_bias_from_cosmic_var, Robertson2014_cosmic_variance_clusters,Chen2023_cosmic_variance_highz, Dalmasso2024_high_bias, Desprez_2024_cv_highz}. Any conclusion about the degree of tension between observations and theory must therefore account for this effect, which is often treated in terms of \textit{cosmic variance}. Cosmic variance is a measure of the degree to which the variance of a distribution of galaxy number counts exceeds the Poisson variance that would be expected if galaxies were distributed randomly. It is therefore also a measure of how much the distribution of galaxy number counts itself differs from the Poisson distribution. \cite{Steinhardt2021} investigated the distribution of number counts of galaxies in the COSMOS2020 \citep{Weaver2022_COSMOS} catalog, and found that a gamma distribution \citep{papoulis2002_probability_book_gamma} provides a good fit to observations.\footnote{Other distributions can be used, as shown by \cite{MBK2010_NegativeBinomial}, but as discussed later, this choice does not affect the results presented in this paper.} The gamma distribution is a natural extension, since it is the analytic continuation of the Poisson distribution in the presence of an additional variance term like cosmic variance.

In this paper, we build upon the work of \cite{Steinhardt2021} to investigate the distribution of the mass of the most massive galaxy in a given field or set of fields under the impact of cosmic variance. We introduce a simple model for computing this distribution, and then compare its predictions with the corresponding distributions in the \texttt{UniverseMachine} simulations, which are based on cosmological N-body simulations. We find good agreement and show several surprising properties of these distributions.

The paper is structured as follows. In \S \ref{sec:methods}, we introduce our simple galaxy model and the gamma distribution, as well as the procedure for finding the most massive galaxy. A brief introduction to the Extreme Value Statistics methodology used in astronomy is also given.
In \S \ref{sec:mostmassive} we present our results and compare them to EVS. The behavior of the distributions of most massive galaxies is analyzed, and the conditions under which cosmic variance needs to be taken into account are shown. The impact of calibrating the cosmic variance to different data sets is also shown.
In \S \ref{sec:simulation}, we validate our assumptions and predictions using the \texttt{UniverseMachine} simulations. As a corollary, we furthermore derive a theorem regarding the variance on empirically calibrated cosmic variance calculator, which may contribute greatly to differences in the predictions of different cosmic variance calculators in the high cosmic variance regime.
In \S \ref{sec:discussion} we discuss how our results impact the field more broadly, and evaluate which aspects of galaxy formation models would currently be most important to constrain. In \S \ref{sec:conclusion} we summarize our conclusions.

\section{Modeling Galaxy Number Counts}
\label{sec:methods}

For the cosmic variance sampling model presented here, a simple model of the stellar mass function at a given redshift $n(M,z)$ is adopted, which can be calculated from an arbitrary halo mass function. Furthermore, a functional form to describe the distribution of number counts is introduced. The functional form is theoretically well-motivated and has already been verified empirically up to $z = 9$ by \cite{Steinhardt2021}. We describe each of the components of the modelling below. In the last subsection of this section, we introduce our base comparison, the Extreme Value Statistics method, as it has traditionally been utilised in astronomy.

\subsection{The Galactic Stellar Mass Function}
\label{sec:mass}

In this work, we connect the dark matter halo mass function (HMF), which can be directly computed for a given cosmology within the $\Lambda$CDM framework, to a galaxy mass function by assuming a one-to-one mapping between galaxy and halo masses at a given redshift. The impact of this choice will be discussed later, but in general, it is negligible in the limit of high cosmic variance. This linking is characterized by the fraction of matter contained in baryons, $f_b$ \citep{Planck2020}, and the stellar baryon fraction (SBF, or $\epsilon_*$), the fraction of baryons contained in stars, such that

\begin{equation}
    \label{eq:ms_mh}
    M_*(z) = f_b \cdot \epsilon_*(z,M_{\mathrm{halo}}) \cdot M_{\mathrm{halo}}
\end{equation} 

\noindent where $\epsilon_*$ is taken as the average value of $\epsilon_*$ for a given bin of halo mass and redshift based on the \texttt{UniverseMachine} \citep{Behroozi2019_UniverseMachine} model, which has been calibrated to match a wide range of observations from $z\sim 0.1$--8. Other possibilities include the model of \cite{Finkelstein2015_SBF}, assuming $\epsilon_*=1$ to get an estimate of the most massive galaxy allowable if using all available baryons, or allowing $\epsilon_*$ to be parameterized as a Gaussian as in \cite{Lovell2023_EVS}.

This one-to-one matching is equivalent to abundance matching \citep{Vale2004_abundancematching}, since the model implicitly assumes a monotonic relationship between galactic stellar mass and halo mass. As noted by \citet{Jespersen2022, Chuang2022, Chuang2023_mangrove}, this ignores a scatter in galactic stellar mass of $\approx 0.3$ dex at fixed halo mass, which could potentially be included in the analytic model, however, in the base model, we choose to not include it to isolate the effect of cosmic variance. However, we stress that the model is flexible enough to allow for any $\epsilon_*(z,M_{\mathrm{halo}})$ relation to be used, monotonic or not.
The model also assumes that all halos host a galaxy, which is not necessarily true \citep{VanDenBosch2007_HaloOccupationStatistics}. The effect of including such a scatter will be shown in \S \ref{sec:discussion}. However, since we here focus on the importance of the inclusion of cosmic variance, this is not the main focus of this paper. The model furthermore assumes a universal baryon fraction, which has been shown to only approximately be true \citep{Crain2007_baryon_frac}. 

For the underlying HMF, the analytic Sheth-Mo-Tormen mass function is assumed \citep{Sheth2001_SMT}. This can also be inaccurate at high redshift, especially for extreme objects, as noted by \cite{Yung23_highz}, who show comparisons between commonly used HMF fitting functions and analytic models with a suite of N-body simulations extending out to $z\sim 20$.
The numerical work with the HMF is done using \verb|HMFcalc| \citep{Murray2013_hmfcalc}.

Given a HMF, and a way of connecting halo masses to galactic stellar masses, we can now obtain a Galactic Stellar Mass Function (GSMF), $n(M_*,z)$. The total number of galaxies can then be obtained by multiplying by the comoving volume that one wishes to consider. In this work, comoving volumes for a given sky area are obtained with \verb|astropy| \citep{Astropy2022}, adopting the cosmology of \cite{Planck2020}, with $H_0 = 67.7 \frac{\mathrm{km}}{\mathrm{Mpc}\cdot \mathrm{s}}$, $\Omega_M = 0.31$ and $\Omega_b = 0.049$. 

\subsection{Cosmic Variance and the Distribution of Galaxy Number Counts}

Cosmic variance is defined in this paper following \cite{Moster2011},

\begin{equation}
    \label{eq:cosmic_variance_defintion}
    \sigma_{\mathrm{CV}}^2=\frac{\left\langle N^2\right\rangle-\langle N\rangle^2-\langle N\rangle}{\langle N\rangle^2}
\end{equation}

\noindent which encodes the extent to which field-to-field variance is higher than expected for a Poisson distribution. The Poisson distribution would be the expected distribution in the absence of clustering and is therefore a natural comparison. The variance is expressed as a fractional variance so that the contribution to the actual variance in a field containing $N$ galaxies is $\sigma = \sigma_{\mathrm{CV}}\cdot N$.

The total variance in a field can therefore be written as the sum of Poisson shot noise and cosmic variance,

\begin{equation}
    \label{eq:total_variance}
    \sigma_{\mathrm{total}}^2 = \langle N^2 \rangle - \langle N \rangle^2= \sigma_{\mathrm{CV}}^2\langle N \rangle^2 + \langle N \rangle
\end{equation}

\begin{figure*}
    \centering
    % [trim={left bottom right top},clip]
    \includegraphics[trim={5cm 2cm -1cm 3cm},clip, width=1.1\textwidth]{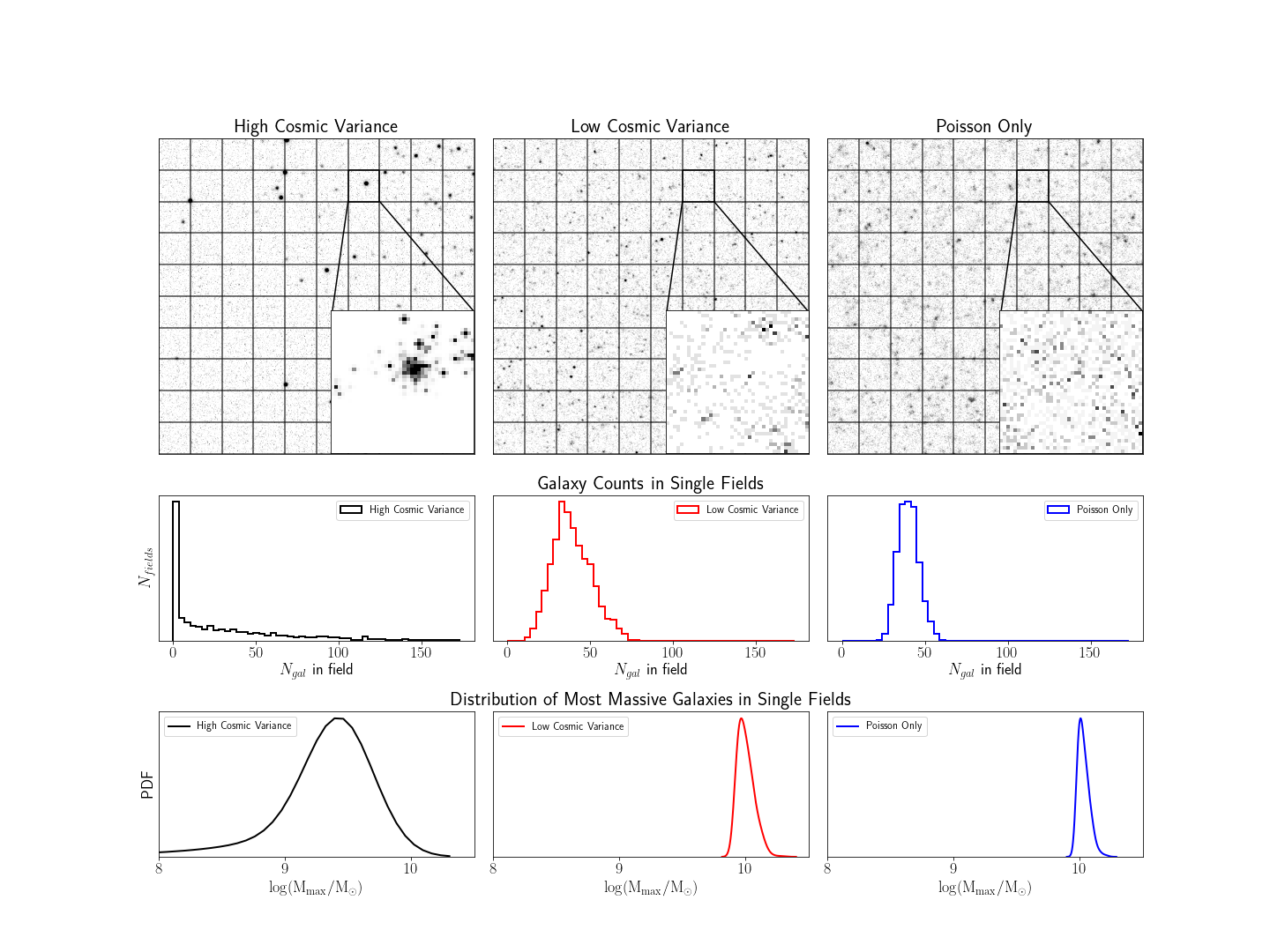}
    \caption{Example simulations of survey fields under different cosmic variance assumptions, visualized with a stellar mass limit of $M_* = 10^8 M_{\odot}$. As clustering and cosmic variance become dominant, the distribution of galaxy number counts changes from approximately Poisson-like to a highly skewed gamma distribution, with a pronounced tail and a high probability of finding no galaxies above the mass threshold. The distribution of most massive galaxies also changes, becoming much wider, and changing from a \textit{positively skewed} (high-mass tail) distribution to a \textit{negatively skewed} (low-mass tail) distribution. The mean and median both become lower and switch places, since for a positively skewed distribution, the median is below the mean due to the high-mass tail, while for a negatively skewed distribution, the median is above the mean due to the low-mass tail. The adopted cosmic variances are $\sigma_{\mathrm{CV}} = \{4, 0.2, 0\}$, the first two being realistic values at $z \geq 10$ for deep JWST surveys, as can be seen in Figure \ref{fig:cv_measurement_UM}.}
    \label{fig:clustering}
\end{figure*}

\noindent Cosmic variance can then be considered an encoding of clustering strength. A visual demonstration of how cosmic variance and clustering are related is given in Figure \ref{fig:clustering}. Since clustering evolves significantly as a function of volume, mass, and redshift, so does cosmic variance.

This work adopts a mass- and redshift-dependent cosmic variance based on the \texttt{UniverseMachine} simulations. Other possible choices could be the cosmic variance calculator of \cite{Moster2011}, which is fitted directly to observations but only at significantly lower redshifts than those considered here, or the simulation-calibrated calculator of \cite{Bhowmick2020}. The cosmic variance calculator of \cite{Moster2011} does not reproduce clustering in \texttt{UniverseMachine} at redshifts significantly outside the calibration range.
However, while the exact choice of cosmic variance calculator impacts the strength of the effects shown in this paper, our qualitative conclusions are independent of the chosen calculator. This is discussed further in \S \ref{sec:mostmassive}. \\

Adding cosmic variance always renders a distribution with the same mean but greater variance than Poisson statistics would predict. To model the distribution of number counts of galaxies, we must therefore choose a distribution which approximates the Poisson distribution in the limit of no extra variance, but with the possibility of setting the variance to be super-Poissonian. The natural choice is the gamma distribution, since it is related to the Poisson but with an arbitrary variance, but other choices are also possible \citep{MBK2010_NegativeBinomial} and will be discussed later. The gamma distribution has been shown to provide a good fit to the distribution of galaxy number counts \citep{Steinhardt2021} and is parametrized as

\begin{equation}
    \label{eq:gamma}
    f(x;k,\theta) = P(X=x) = \frac{x^{k-1}e^{-\frac{x}{\theta}}}{\theta^k\Gamma(k)}
\end{equation}

\noindent where $\Gamma$ is the Euler gamma function, and $k$ and $\theta$ are the shape and scale parameters. The shape and scale are useful since they allow us to independently set the mean ($\mu = k\theta$) and variance ($\sigma^2 = k\theta^2$) of the distribution.\footnote{Where $\mu$ is the expected/average number of galaxies.}  We therefore have  

\begin{eqnarray}
\label{ktheta}
    k & = & \left(\frac{\mu}{\sigma_{\mathrm{total}}}\right)^2  \nonumber \\
    \theta & = & \frac{\sigma_{\mathrm{total}}^2}{\mu}
\end{eqnarray}

\subsection{Estimating the Maximum Value Distribution}
\label{subsec:max_val_dist_procedure}

To model the effects of clustering, we sample each field by assuming that if a field is an outlier in a given 0.5 dex mass bin, it will be an outlier to the same degree for galaxies of all masses. For example, if the number of $10^{8.5} M_{\odot}$ galaxies is an $n \sigma$ outlier for a given field, all other numbers are drawn such that the number of galaxies for that mass is also an $n \sigma$ outlier. This assumption implies that the number counts of galaxies of all masses are fully correlated, which is obviously not true, but turns out to be a reasonable approximation, especially for smaller fields, where the same long-wavelength density modes dominate. This mass scale covariance can be clearly observed in the recent FLARES simulations \citep{FLARES-I, FLARES-II}, a series of zoom simulations of a range of overdensities at high redshift ($z > 5$); Figure 9 of \citealt{Thomas2023_FLARES_overdensity} shows the GSMF as a function of overdensity. A similar correlation has also been found in recent observations of high-z massive, quiescent galaxies by \cite{Valentino23_quiescentAtlas}, although the mass range considered here was limited to $9 < \mathrm{log}(M_*/M_{\odot}) < 12$. The validity of this assumption also relies on the width of the adopted mass bins, since the relation would be much noisier for small bins. However, for the 0.5 dex mass bins adopted here, the noise is generally not an issue. The impact of assuming fully correlated mass scales will be discussed further in \S \ref{sec:discussion}. The exact magnitude of the correlations will be discussed further in Appendix \ref{appsec:masscales}.

Sampled galaxies are compared and the most massive galaxy of that sample is found. The cumulative distribution function (CDF) of the distribution of most massive galaxies is then obtained by sampling the distribution in rolling mass bins\footnote{This is also often referred to as a sliding box-car, meaning that each bin can overlap with other bins. For example, the bin centered at $M = 10^{8.1}$ includes the interval $M = 10^{7.85}-10^{8.35}$, but we also include other bins which cover much of the same interval, like $M = 10^{8}$ which includes the interval $M = 10^{7.75}-10^{8.25}$.} of 0.5 dex width and combining the CDFs of most massive objects to make a high-resolution CDF. This ``supersampling'' approach makes the resulting CDF invariant to the exact mass bins used. The probability distribution function (PDF) of the mass of the most massive galaxy can then be obtained by taking the derivative of the CDF. Since our approach relies on sampling objects from a gamma distribution with properly added cosmic variance, we will use the term \textit{cosmic variance sampling model} to refer to the model presented in this paper.

To summarize the procedure in a few steps:
\begin{enumerate}
    \item For a given redshift ($z$), redshift bin size ($dz$), and survey design (characterized by side lengths $s_1,s_2$ and total area $A$), compute the comoving volume, $V(z, dz, A)$, and obtain the absolute $\mathrm{HMF}$ as $V\cdot \mathrm{HMF}(z)$.
    \item Translate $\mathrm{HMF}(z)$ to $\mathrm{SMF}(z)$ as described in Eq. \ref{eq:ms_mh}, and integrate $\mathrm{SMF}(z)$ in mass bins centered at $M$ and with a width of 0.5 dex to obtain $\langle N_M \rangle$, the \textit{mean} number of galaxies in that mass bin.
    \item For each mass bin, define the gamma distribution $\Gamma_M$ (Eq. \ref{eq:gamma}) with $\sigma_{\mathrm{total,M}}^2 = (\sigma_{\mathrm{CV}}(z, dz, M, s_1, s_2) \langle N_M \rangle)^2 + \langle N_M \rangle $ and $\mu = \langle N_M \rangle$. This is the only step that would be affected by changes in the cosmic variance.
    \item Fairly sample the ``clustering percentile'', $u$. $u$ will be in the $[0, 1]$ interval, where e.g., a value of 0.5 correspond to the \textit{median} field, 0.84 to a +1$\sigma$ \textit{overdensity}, and 0.16 to a 1$\sigma$ \textit{underdensity}. 0 would correspond to the most underdense field in the universe, and 1 would correspond to the most overdense field in the universe. The same $u$ is used for all mass bins in that field. For each mass bin and the corresponding $\Gamma_M$, we can then invert the CDF at $u$ to give an observed number of galaxies $\widehat{N_M}$. The highest $M$ with $\widehat{N_M} \geq 1$ is taken as the mass of the most massive galaxy for that sample.
    \item This is then repeated for a large set\footnote{In practice this is done by taking $10^5$ linearly spaced points in this interval.} of $u$-values in the $[0, 1]$ interval to give a fully sampled distribution of the masses of the most massive galaxies in single fields at a given $z$, $dz$, and survey design. Repeat the process for different redshifts to get the redshift evolution.
\end{enumerate}

The width of the resulting distribution of the masses of the most massive galaxies, $\sigma_{\mathrm{M}}$ will be dominated by different terms. The width due solely to a given component x is then $\sigma_{\mathrm{M,x}}$, so the width due to the inclusion of cosmic variance would be $\sigma_{\mathrm{M,CV}}$, while the width due to Poisson noise only would be $\sigma_{\mathrm{M,Poisson}}$. These terms will be evaluated later on in order to make a comparison with the width due to scatter in the stellar-to-halo-mass relation, $\sigma_{\mathrm{M,M_h-M_*}}$, which has been proposed frequently as a possible solution to the tension between observations and galaxy formation models \citep{Dekel2023_bursts, Shen2023_UV_burst_highz, Endsley_2024_bursty_highz}.

\subsection{Extreme Value Statistics}
\label{subsec:EVS}

The main reference for comparison of our method is the Extreme Value Statistics (EVS) technique \citep{gumbel_statistics_1958,katz_extreme_2000}, which predicts the greatest or lowest valued random variable from a given distribution.
EVS has been applied in a number of areas of astronomy, including the mass of the most massive local galaxy clusters \citep{harrison_exact_2011,harrison_testing_2012,harrison_consistent_2013,chongchitnan_primordial_2012,waizmann_application_2012, Busillo2023_EVS} as well as the sizes of voids \citep{chongchitnan_abundance_2015,sahlen_clustervoid_2016}.
\cite{Lovell2023_EVS} applied EVS to the prediction of the most massive haloes and galaxies in the early Universe.

The EVS approach is summarised as follows.
Consider a cumulative distribution function (CDF), $F(m)$; if we draw $\{N\}$ random variates $\{M_i\}$ from this distribution, assuming all variables are mutually independent and identically distributed (IID), there will be a largest value of the sequence, $M_{\mathrm{max}} \equiv \mathrm{sup}\{M_1 \dots M_2\}$, and the probability all draws are less than some value $m$ is given by 
\begin{eqnarray}
\Phi (M_{\mathrm{max}} \leqslant m;\, N) & = & F_{1} (M_{1} \leqslant  m) \; ... \; F_{N} (M_{N} \leqslant  m) \\
& = & F^{N} (M)
\end{eqnarray}
To obtain the probability density function (PDF) of the distribution we differentiate,
\begin{eqnarray}
\Phi (M_{\mathrm{max}} = m;\, N) & = & N F'(m) [ F(m) ]^{N-1} \\
& = & N f(m) [ F(m) ]^{N-1} \;.
\end{eqnarray}
Here, $f(m)$ is the PDF of the original distribution ($f(m) = dF(m)/dm$), and $\Phi (M_{\mathrm{max}} = m;\, N)$ is the \textit{exact} extreme value PDF for $N$ observations drawn from the known CDF, $F(m)$, given that the above IID assumptions hold.

The PDF and CDF for haloes in a fixed fraction of the sky, $f_{\mathrm{sky}}$, between redshifts $z_{\mathrm{min}}$ and $z_{\mathrm{max}}$ is given by:
\begin{eqnarray}
f(m) & = & \frac{f_{\mathrm{sky}}}{n_{\mathrm{tot}}} \left [ \int_{z_{\mathrm{min}}}^{z_{\mathrm{max}}} dz \frac{dV}{dz} \frac{dn(m,z)}{dm} \right ] \\
F(m) & = & \frac{f_{\mathrm{sky}}}{n_{\mathrm{tot}}} \left [ \int_{z_{\mathrm{min}}}^{z_{\mathrm{max}}} \int^{m}_{- \infty}  dz \, dM \frac{dV}{dz} \frac{dn(M,z)}{dM} \right ] \;,
\end{eqnarray}
where
\begin{equation}
n_{\mathrm{tot}} = f_{\mathrm{sky}} \left [ \int_{z_{\mathrm{min}}}^{z_{\mathrm{max}}} \int^{\infty}_{- \infty}  dz \, dM \frac{dV}{dz} \frac{dn(M,z)}{dM} \right ] \;,
\end{equation}
and $dn(m,z)\,/\,dm$ is the assumed halo mass function at redshift $z$.
Together these can be used to find the halo EVS for a given survey, and combined with similar assumptions for the baryonic and stellar fractions as explained above, the stellar mass of the most massive galaxy. As emphasised previously, this form of EVS cannot take into account the impact of cosmic variance, since EVS relies on the IID assumption, which by its very nature cannot include clustering/cosmic variance effects.

\section{The Most Massive Objects in \textit{JWST} Fields}
\label{sec:mostmassive}

\begin{figure*}
    \centering
    \includegraphics[trim={0cm 0cm 0cm 0cm},clip, width=0.77\textwidth]{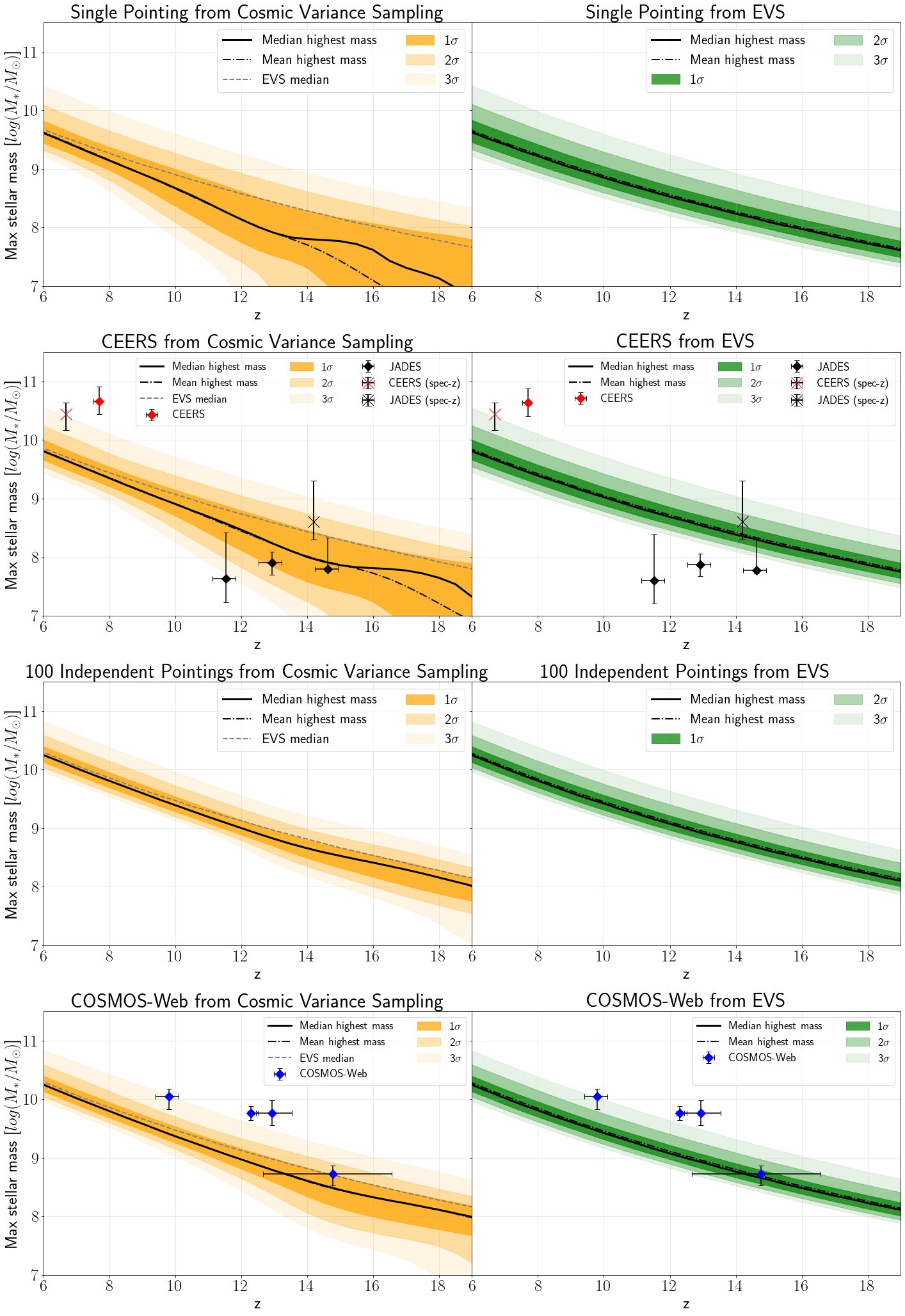}
    \caption{The predicted distributions of the mass of the single most massive object in a survey as a function of redshift in four surveys: A single \textit{JWST} pointing of 9.7 arcmin$^2$; a medium-deep CEERS/JADES-like survey of 35 arcmin$^2$ 100 entirely independent \textit{JWST} pointings, such as in a snapshot or parallel survey, with a total area of 0.27 deg$^2$; and a wider survey similar to the COSMOS-Web Survey, of 0.28 deg$^2$. Predicted distributions from the cosmic variance sampling model are shown in the left column, while predictions from EVS, which does not take into account cosmic variance, are shown in the right column. The two approaches yield significantly different results, especially for high-redshift, small area surveys. Calculations are done for a running redshift interval of $\Delta z = 1$. The dashed and solid lines indicate the mean and median of the distribution, and contours indicate 1, 2, and 3-$\sigma$ intervals. If possible, the most massive, high-redshift galaxies detected for the relevant corresponding \textit{JWST}-survey are shown. References to the objects are given in the text. Only galaxies with either spectroscopic redshifts or non-parametric fits to multi-band photometry are considered. We also show the limit for any significant ($2 \sigma$) tension with $\Lambda\text{CDM}$ for both approaches. This is evaluated by running each model with 100$\%$ star formation efficiency ($M_*=M_{\mathrm{baryon}}$) and taking the corresponding upper $2 \sigma$ contour. It is clear that these galaxies do not imply cosmological tension.}
    \label{fig:singleobject}
\end{figure*}
 
The impact of cosmic variance on the distribution of most massive objects is found by comparing the difference between the cosmic variance sampling model and the EVS framework. The probability distribution of these extreme galaxies is considered for four possible survey designs:

\begin{itemize}

    \item A single \textit{JWST} NIRCam pointing of 9.7 arcmin$^2$.  Although wider surveys will likely be more common, single pointings may comprise the deepest available observations, especially for lensing cluster fields \citep[e.g.][]{Treu2022_GLASS}. This is similar to UNCOVER, although here, the effects of lensing have to be carefully modelled, which is why no objects are shown in the corresponding panel \citep{Vujeva2023_lensing}.
    
    \item A contiguous survey area of 35 arcmin$^2$ similar in design to the bf CEERS or JADES field \citep{Finkelstein2017_CEERS, Williams2018_JWST_JADES_forecast}. The objects shown in the corresponding panel in Figure \ref{fig:singleobject} are CEERS-44057\footnote{Spectroscopically confirmed.}, CEERS-9317, JADES+53.09731-27.84714, JADES+53.06475-27.89024, JADES+53.10762-27.86013, taken from \citet{Chworowsky2023_Ceers} and \citet{Robertson2023_JADES_results}. JADES-GS-z14-0\footnote{Spectroscopically confirmed.} is also included, with the mass taken from the original detection paper by \cite{Carniani2024_JADES-GS-z14-0}, whereas the redshift is taken from the much more precise ALMA observation of the same source \citep{Schouws2024_JADES-GS-z14-0_ALMA}.
    
    \item A snapshot survey consisting of bf 100 independent pointings. This could be done as an archival study of background objects in early \textit{JWST} programs, as a targeted program or as a parallel program, similar in design to HST-BoRG \citep{Trenti2012_HST_Borg, Holwerda2019_parallel}.\footnote{This would be the most efficient way of reducing the impact of cosmic variance, since for N completely uncorrelated fields $\sigma_{\mathrm{CV,total}} = \sigma_{\mathrm{CV,field}}/\sqrt{N}$.} This covers a total area of $\approx 0.27 \mathrm{deg}^2$, which is similar to the area that was surveyed to find the pre-\textit{JWST} highest redshift galaxy, GN-z11 \citep{Oesch2016}.

    \item A wider area single continuous survey similar to the COSMOS-Web survey \citep{Casey2023_CosmosWebDesign}. This survey will cover different areas depending on the imaging filter. Here we adopt an effective area of 0.28 deg$^2$, which is representative of the area covered at the time of writing of this paper, and also makes for easy comparison with the above snapshot survey. The objects shown in the corresponding panel of Figure \ref{fig:singleobject} are COS-z10-2, COS-z12-3, COS-z13-2, COS-z14-1, and are taken from \cite{Casey2023_CosmosWebResults}. 
\end{itemize}

For the considered surveys, we compare our results to the results from the EVS framework, adjusted to also correspond to 0.5 dex mass bins, a redshift window of $\Delta z=1$, and the same cosmology, HMF and halo-to-stellar mass conversion model.

The probability distribution for the stellar mass of most massive objects as a function of redshift from the two approaches are shown in Figure \ref{fig:singleobject}. Figure \ref{fig:singleobject} also shows current massive galaxy estimates in the relevant fields. Only galaxies with spectroscopy or non-parametric fits to multi-band photometry are considered, as we deem others to be too uncertain to be included. Since the possibility of the observed galaxies requiring changes to $\Lambda$CDM has been brought up in the literature, we also show the upper 2$\sigma$ limit for models run with $\epsilon_* = 1$, which is a representative limit for absolute tension with $\Lambda\text{CDM}$ \citep{Lovell2023_EVS, Steinhardt2023_BalmerBreaks}.

As can be seen in Figure \ref{fig:singleobject} , the two approaches agree qualitatively in the limit of low cosmic variance (low redshift and large survey area). However, as the impact of cosmic variance grows (for smaller areas or higher redshifts) the distributions diverge rapidly. In the small field/high-z limit (where the cosmic variance is high), the distributions from the cosmic variance sampling model show a significant low-mass skew (mean below median) which is in the opposite direction from the slight high-mass skew (mean above median) of the EVS and low cosmic variance distributions. 
There are now two possible ways of evaluating the tension between the data and the models shown here. The first is the global predictive ability, where we consider the models' abilities to accurately predict the masses of all observed massive galaxies. In Figure \ref{fig:singleobject}, we see that the cosmic variance sampling provides an overall better fit to most of the observed data, especially for the smaller fields. However, since the main tension with current galaxy formation models has been the observation of too many massive galaxies \citep{Mirocha2023_highz}, we may also wish to evaluate the one-sided tension with each model. This test is done since one can theoretically make high redshift galaxies have lower mass by invoking a wide array of mechanisms, but making them more massive requires some re-evaluation of our models \citep{Dekel2023_bursts}. The one-sided tension can be evaluated using the statistical method called using Fisher's Method \citep{heard2018_pvalues}. For example, when combining the likelihoods of all of the CEERS/JADES galaxies (2nd row in Figure \ref{fig:singleobject}) using Fisher's Method, the total p-value without cosmic variance is $p = 0.075$ (corresponding to $\sigma_{\mathrm{all}} \approx 1.4$), but with cosmic variance, it is $p = 0.0035$ (corresponding to $\sigma_{\mathrm{all}} \approx 2.7$). Similarly, for COSMOS-Web, the total p-value goes from $p = 0.002$ ($\sigma_{\mathrm{all}} \approx 2.9$) when using EVS to $p = 0.0003$ ($\sigma_{\mathrm{all}} \approx 3.4$) when considering cosmic variance. The inclusion of cosmic variance thus \textbf{increases} the tension by almost $1\sigma$ for small fields, and $0.5\sigma$ for large fields. This is mainly based on photometric objects, but the spectroscopic objects still represent some tension with galaxy models. However, the observed galaxies do not imply any significant tension with $\Lambda\text{CDM}$, as can be seen by the red dashed line in Figure \ref{fig:singleobject}, which shows each respective model run with 100$\%$ star formation efficiency.

The shift in the shape of the distribution of most massive galaxies can be physically understood by considering that high cosmic variance implies high clustering. Then, any field where a single massive galaxy is found is also highly likely to be populated by a host of galaxies of lower mass. Conversely, most fields will be almost empty, except for a few very low-mass galaxies (which will be below the detection limit of \textit{JWST}). Because of this behavior, the distribution of the mass of the most massive galaxies will have a low-mass skew, and the mean of our distribution will lay significantly below the median. This is in contrast to the distribution of number counts, where the skew always goes towards higher numbers (positive) when increasing the cosmic variance/clustering. However, the effect on the most massive galaxy is the opposite.

\begin{figure}
    \centering
    \includegraphics[trim={0cm 0.cm 0.cm 0.cm},clip,width=1.\linewidth]{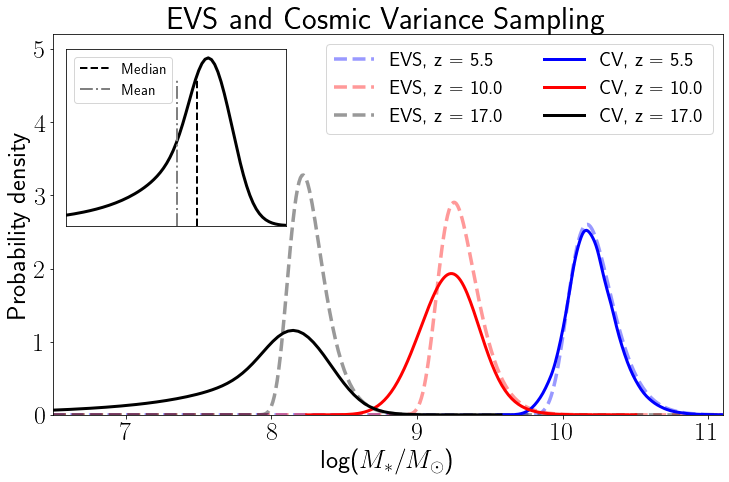}
    \caption{Probability distribution of the mass of the highest mass galaxy in a given field from the cosmic variance sampling model and EVS, at three different redshifts. The predicted distribution in the low cosmic variance limit from the cosmic variance sampling model agrees almost perfectly with the EVS predictions. However, in the high-z limit, as field-to-field variance starts to dominate, the distributions diverge significantly, and the hierarchy of the mean and median is reversed. The exact redshift where this divergence takes place is highly dependent on the survey geometry and redshift range considered, since these factors change the cosmic variance.}
    \label{fig:EVScomparison}
\end{figure}

This means that when including cosmic variance, the distribution of the masses of the most massive galaxies will lie below those calculated with EVS.
The two distributions from the two approaches converge at the highest masses, since the predicted highest-mass galaxies in the universe will be the same in both approaches. This is shown explicitly in Figure \ref{fig:EVScomparison}, which shows the predicted distributions of most massive galaxies as a function of mass only. The difference arises since, when taking into account field-to-field clustering, the remaining galaxies simply tend to cluster in the same fields, but since only the single most massive galaxy in a field is considered, the other galaxies carry no impact in that field.

Since cosmic variance is usually only invoked for number count distributions, where the skew always becomes increasingly positive as the cosmic variance goes up, these results may seem contrary to prior expectations. While the shift from the expected positive (high-mass) skew of the distribution to a negative (low-mass) skew in the high cosmic variance limit could therefore be seen as counterintuitive, it can easily be visualized as an effect of clustering, as shown in Figure \ref{fig:clustering}. There is also further theoretical basis for this results, as it is a well-known case in Extreme Value Theory \citep{mises1936_EVT}. Extreme Value Theory is not to be confused with the EVS method as applied to astronomy, as Extreme Value Theory is a much broader field.
All possible distributions of sampling extreme values can be described through the Generalized Extreme Value Distribution \citep{FisherTippett_1928, gnedenko1943distribution}, for which there are three separate limits. The Reversed Weibull (or Type III Extreme Value Distribution) limit is the relevant limit for the high cosmic variance, negatively skewed distribution. This limit is relevant when significant additional variance is present in the underlying distribution of galaxy counts, regardless of whether all draws are taken to be correlated or not.\footnote{Correlating draws will, however, impact the magnitude of the change in the skew of the distribution.} This is consistent with the findings of \cite{Harrison2011_EVS}, who showed that the positively skewed Type I Extreme Value Distribution is inconsistent with the distribution of the masses of most massive halos.

While not investigated by those authors, the negatively skewed limiting case is also identified in the EVS approach of \cite{Lovell2023_EVS}. When EVS is combined with drawing $\epsilon_*$ from some underlying PDF with high variance, the predicted distribution of the masses of the most massive galaxies shifts from positively to negatively skewed, mimicking the impact of cosmic variance (see Figure 3 of \cite{Lovell2023_EVS}). Drawing a field-dependent stochastic $\epsilon_*$ corresponds to varying all mass scales similarly to that presented in this paper, so the fact that similar effects are seen is unsurprising. 

\begin{figure*}[h!]
    \centering
        % [trim={left bottom right top},clip]
    \includegraphics[trim={0cm 0cm 0cm 0cm},clip,width=1.05\linewidth]{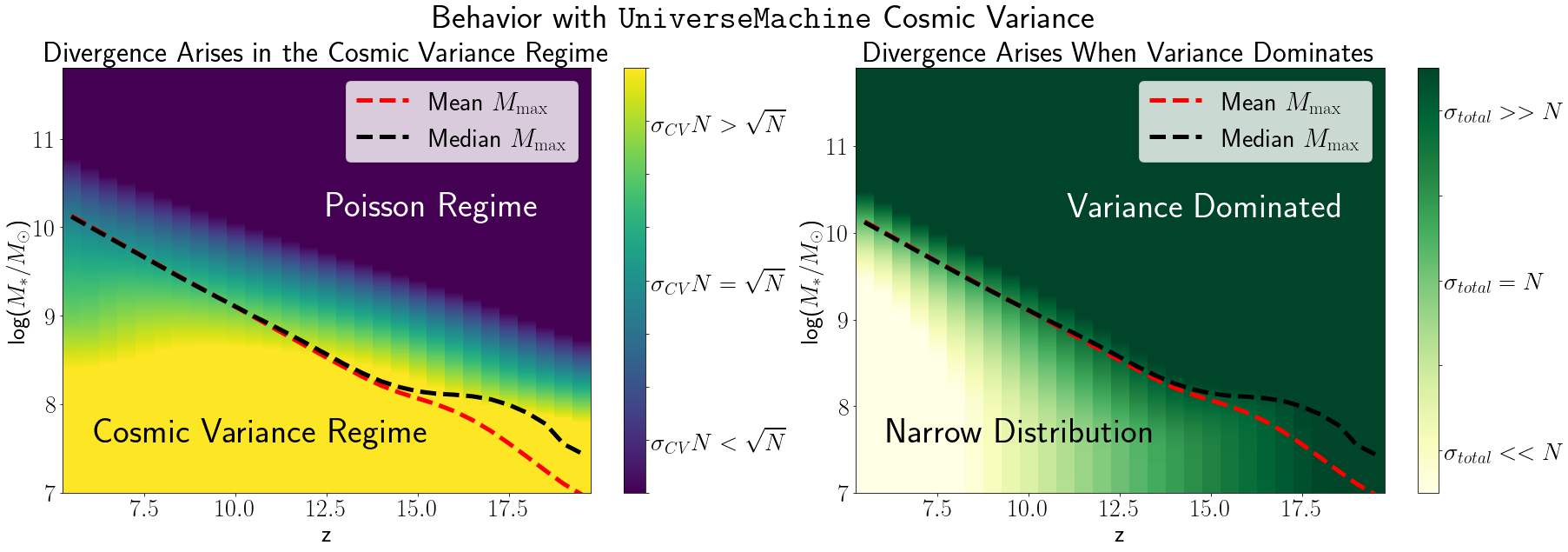}
    \caption{Mean and median of the distribution of the masses of the most massive galaxies, calculated using the cosmic variance from \texttt{UniverseMachine}. The slight positive (high-mass) skew of the distribution of the mass of the most massive galaxy in the low cosmic variance limit is replaced by a negative (low-mass) skew when the cosmic variance becomes larger than the Poisson shot noise, i.e. when the variance is dominated by clustering ($\sigma_{\mathrm{CV}}\langle N \rangle > \sqrt{\langle N \rangle}$, with N being the number of galaxies, as shown in the left panel) and the skew causes a divergence between the mean and median when the total variance becomes larger than the average number of galaxies in the field, i.e. becomes variance dominated ($\sigma_{\mathrm{CV}}\langle N \rangle > \sqrt{\langle N \rangle}$, as shown in the right panel). The values shown here are calculated for a CEERS-like survey.} 
    \label{fig:poisson_cv}
\end{figure*}

It is known that the distribution of maximum values is positively skewed (median below mean) when the variance of the underlying number count distribution is mostly Poisson-like ($\sigma_{\mathrm{total}} \approx \sqrt{\langle N \rangle}$ in our case, with N being the number of galaxies) and the number count distribution is not heavily skewed, i.e. $\sigma_{\mathrm{total}}<<N$ \citep{haan2006_ExtremeValueTheory_book}. As described by \cite{haan2006_ExtremeValueTheory_book}, it is also expected that the distribution of maximum values is negatively skewed (mean below median) in the limit where the variance dominates, regardless of origin. In our case, this limit is reached with a large contribution from cosmic variance ($\sigma_{\mathrm{total}} \approx \sigma_{\mathrm{CV}}\langle N \rangle$ and $\sigma_{\mathrm{total}}>>N$). This is exactly what is observed in Figure \ref{fig:poisson_cv}, where the inversion of mean and median takes place exactly when the variance starts dominating the number count distribution, due to the increased contributions from cosmic variance.

This behavior is independent of the underlying distribution assumed \textbf{if} the variance of the number count distribution from which we are sampling can be set to be super-Poissonian. However, this is a necessary condition for any distribution describing the number counts of galaxies in all regimes. A different distribution which has been suggested to possibly describe the number counts of galaxies is the \textit{negative binomial} distribution \citep{MBK2010_NegativeBinomial, deSouza2015, HurtadoGil2017_NegativeBinomial, Perez2021_NegativeBinomial}. However, parameterizing the distribution of galaxy number counts using the negative binomial does not result in any substantial change in the behavior of the distribution of most massive galaxies. The results presented in this paper are therefore robust to any reasonable choice of the distribution of galaxy number counts. 
The appropriateness of either of these distributions is discussed further in \S \ref{sec:distribution}.

\begin{figure*}[h!]
    \centering
    \includegraphics[trim={0cm 0cm 0cm 0cm},clip, width=1.05\linewidth]{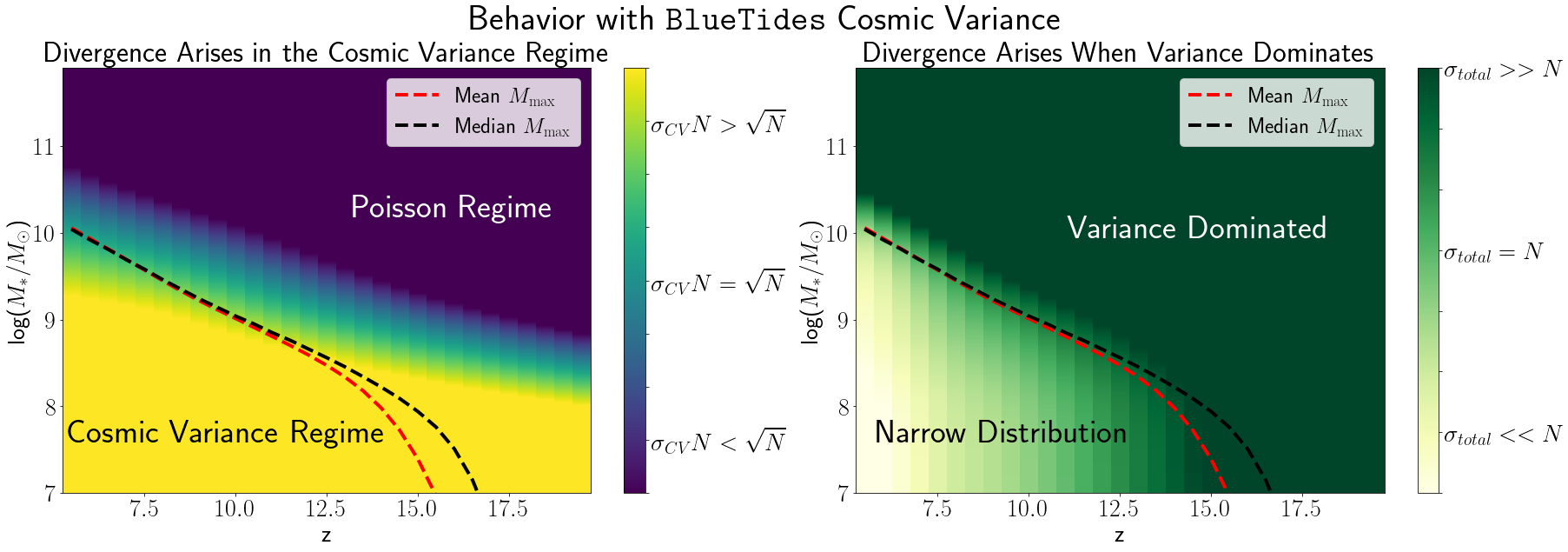}
    \caption{Same as Figure \ref{fig:poisson_cv}, but using cosmic variance calculated based on the \texttt{BlueTides} cosmic variance. (Left panel) The onset of the negative skew in the distribution of the mass of the most massive galaxy takes place at a slightly lower redshift when using the \texttt{BlueTides} cosmic variance estimates, but it happens according to the same condition ($\sigma_{\mathrm{CV}}\langle N \rangle > \sqrt{\langle N \rangle}$) as for the \texttt{UniverseMachine} cosmic variance. (Right panel) The divergence of the mean and median happens at a lower redshift, but much more rapidly. The condition is the same as before, i.e. that the underlying distribution of galaxy number counts is variance dominated ($\sigma_{\mathrm{total}} > \langle N \rangle$). Both conditions are therefore stable, but take place at different z, with different magnitudes. For example, compared to the above figure, the mass of the mean most massive galaxy at $z = 12$ is a quarter dex higher, while at $z = 15$, it is a dex lower} 
    \label{fig:bluetides_noise_poisson_cv}
\end{figure*}

Because the transitions are dependent on the redshift and mass at which $\sigma_{\mathrm{CV}}\langle N \rangle > \sqrt{\langle N \rangle}$ and $\sigma_{\mathrm{total}} = \sqrt{\sigma_{\mathrm{CV}}^2\langle N \rangle^2 + \langle N \rangle} >> \langle N \rangle$, they are necessarily highly dependent on the way that the cosmic variance is calculated. Many calculators are available, and render very different predictions \citep{TrentiStiavelli2008_CV_Calculator, Moster2011, Bhowmick2020, Trapp2020_cosmicvariance_calculator, Ucci2021_cosmicvariance_calculator}. Here we test the impact on our results of implementing the calculator calibrated on the \texttt{BlueTides} simulation \citep{Bhowmick2020}. The \texttt{BlueTides} simulations were specifically made for investigating galaxy formation at high redshift \citep{Feng2016_BlueTides_intro}. The \texttt{BlueTides} calculator gives cosmic variance in terms of UV magnitudes, which are converted to masses by using the $M_* - M_{\mathrm{UV}}$ - relation from \cite{song2016evolution}. The results from using the cosmic variance from \texttt{BlueTides} are shown in Figure \ref{fig:bluetides_noise_poisson_cv}.

Doing the same calculation with the \texttt{BlueTides} calculator (Figure \ref{fig:bluetides_noise_poisson_cv}), we observe that the transition between a positively and negatively skewed distribution of massive galaxies, and subsequent inversion and divergence of mean and median take place at different redshifts. However, the overall behavior of the distribution of the masses of the most massive galaxies is the same, with mean and median crossing when cosmic variance is higher than Poisson variance, and the divergence of mean and median taking place when the total field-to-field variance is higher than the average number of galaxies per field. The qualitative behavior predicted in this paper is therefore robust to the choice of cosmic variance calculator. The calculator of \cite{Moster2011} was also tested, yielding the same qualitative behaviour.

\section{Validation in Cosmological Simulations}
\label{sec:simulation}

Since cosmic variance is usually thought of as adding extra positive skew \citep{Steinhardt2021}, these results may seem contrary to prior expectation, and the assumptions and consequences of the model must therefore be validated in simulations. The validations are done using the publicly available catalogs from \texttt{UniverseMachine}\footnote{\url{https://halos.as.arizona.edu/UniverseMachine/DR1/JWST_Lightcones/}} \citep{Behroozi2019_UniverseMachine} since \texttt{UniverseMachine} is tuned directly to reproduce observed galaxy clustering. 32 \texttt{UniverseMachine} lightcones are used, eight from each of four different survey emulations, \texttt{COSMOS} (17'x41'), \texttt{UDS} (36'x35'), \texttt{GOODS-N} (69'x32') and \texttt{GOODS-S} (45'x41'). Each of the four survey designs then contains eight realizations of different initial conditions. All lightcones extend to a redshift of 17.  We validate the choice of the gamma function as the parametrization of the distribution of galaxy number counts, the onset of negative skew in the distribution of the mass of the most massive galaxy in a given field as a consequence of high cosmic variance and that the distribution of the mass of the most massive objects follows the evolution with redshift predicted by the model in this paper (shown in Figure \ref{fig:singleobject}).\\

As a further point of investigation, we present a theorem on the spread in measured\footnote{The reader should note that this spread is solely a statistical spread.} cosmic variances, and empirically demonstrate its validity. Compared to the assumptions of previous work on cosmic variance calculators which traditionally assume uniform Gaussian errors \citep{Moster2011, Bhowmick2020}, the spread in measured cosmic variance at a given redshift is order of magnitudes larger in the limit of high cosmic variance. This may explain some of the divergence between the predictions of different cosmic variance calculators, and should be included in future calculator calibrations. 

\subsection{The Distribution of Galaxy Counts}
\label{sec:distribution}

As discussed in \S \ref{sec:mostmassive}, different parametrizations of the distribution of galaxy number counts are possible. To verify the choice of the gamma distribution made in this paper, a wide range of distributions are tested. All 103 distributions which allow for fitting in the \texttt{scipy} library were tested, using four different metrics, for sampled 4'x4' fields from \texttt{UniverseMachine}. The applied metrics are the $\chi^2$ \citep{barlow1993_statistics}, the Kolmogorov-Smirnov (KS) - test \citep{kolmogorov1933_kstest}, and the Anderson-Darling test (AD) \citep{anderson1952asymptotic}. Although some fit better in a few cases, the gamma and negative binomial distributions are globally preferred across the different \texttt{UniverseMachine} boxes and the different metrics. This should not be surprising, as these fulfill the conditions discussed by \cite{MBK2010_NegativeBinomial} and \cite{Steinhardt2021} for any distribution describing galaxy number counts, both being continuations of the Poisson distribution with the possibility of increasing the variance separately with an additional term.

\begin{figure}
    \centering
    \includegraphics[trim={0cm 0.cm 0.cm 0.1cm},clip,width=0.9\linewidth]{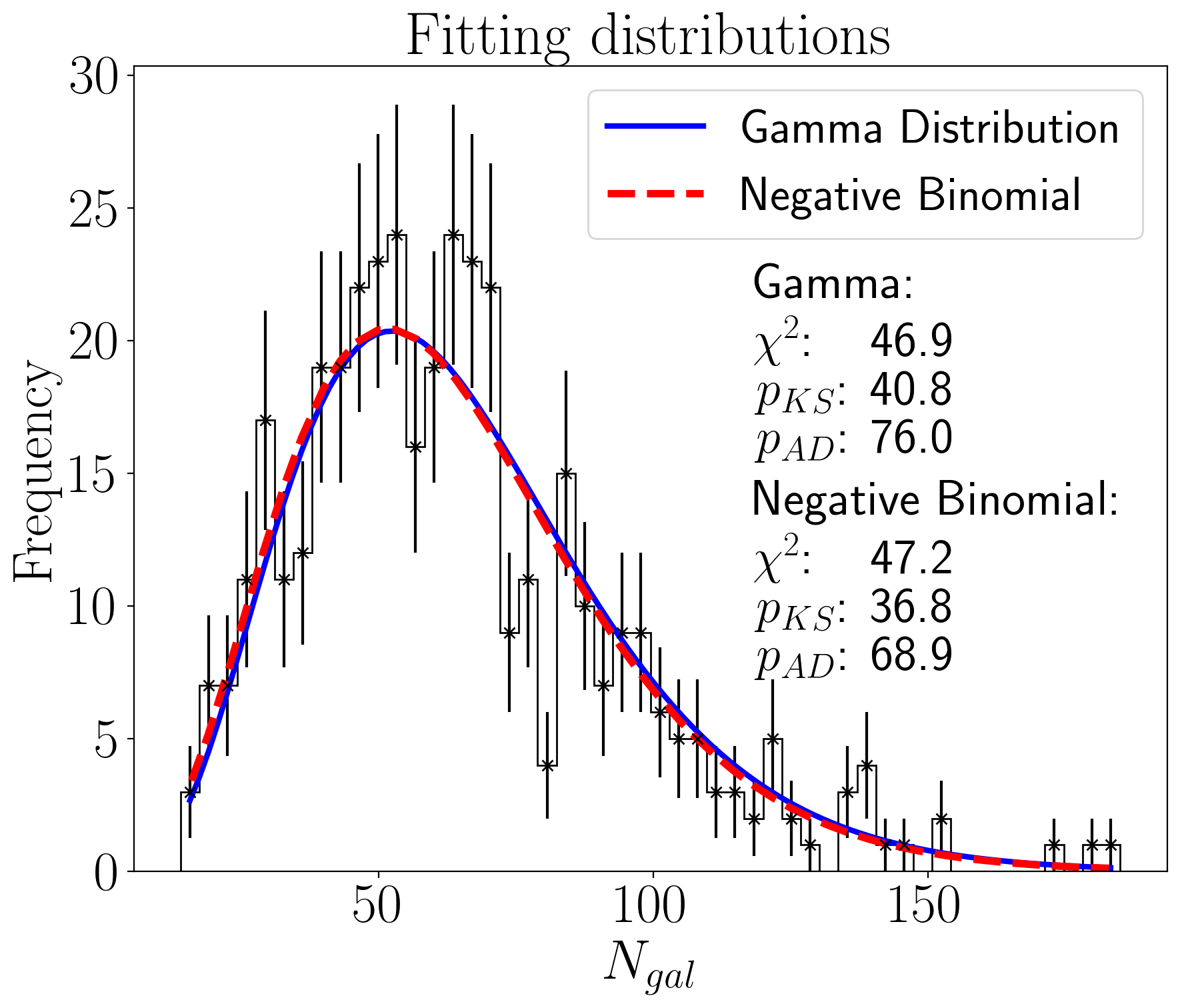}
    \caption{An example of fitting the gamma and negative binomial to sampled galaxy number counts from \texttt{UniverseMachine}. Although in this case, there is a slight preference for the gamma distribution, the difference is negligible. The use of both distributions is therefore valid.}
    \label{fig:fitting_example}
\end{figure}

We therefore focus on evaluating the goodness of fit of these two distributions, the results of which can be seen in Table \ref{tab:nb_gamma_comparison}. An example fit can be seen in Figure \ref{fig:fitting_example}. There is no clear preference for either distribution, so a statistical argument can be made for choosing either. Here, the gamma is preferred, since this distribution fits the tail of the galaxy count distribution better than the negative binomial, as indicated by the Anderson-Darling test.\footnote{However, the negative binomial natively supports integer values, which may be helpful for some applications.}

\begin{table}
    \centering
    \caption{Preferred distribution for each metric for the two globally best-fitting distributions, the gamma distribution and the negative binomial (NB), to the distribution of galaxy number counts in \texttt{UniverseMachine} for three different field sizes. Both distributions are fitted to have the same mean and variance as the number count distribution. There is no clear preference. The metrics considered are the $\chi^2$, the Kolmogorov-Smirnov (KS) - test, and the Anderson-Darling test (AD).}
    \label{tab:nb_gamma_comparison}
    \begin{tabular}{|l|c|c|c|c|c|}
    \hline
     Field sizes & $\chi^2$ & KS & AD \\
    \hline
     2'x2' & Gamma & NB & Gamma\\
    \hline
    4'x4' & NB & Gamma & Gamma\\
    \hline
    8'x8' & NB & Gamma & Gamma\\
    \hline
    \end{tabular}
\end{table}

As mentioned in \S \ref{sec:mostmassive}, the impact of choosing either the negative binomial or the gamma distribution to parametrize the distribution of galaxy number counts is negligible. This is unsurprising given that they both provide similarly good fits. 

\subsection{The Skew of the Distribution of Most Massive Objects}

\begin{figure*}
    \centering
    % [trim={left bottom right top},clip]
    \includegraphics[trim={0cm 0cm 0cm 0.cm},clip,width = .9\linewidth]{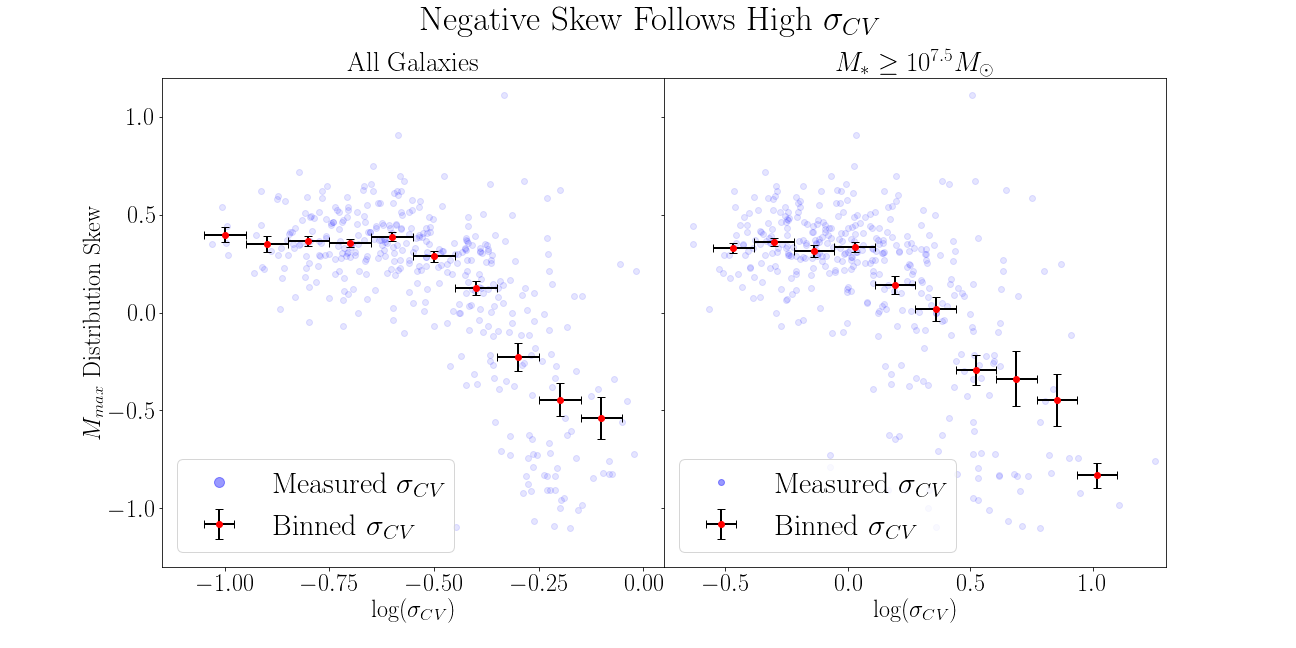}
    \caption{The scaling of the skew of the distribution of the mass of the most massive galaxy as a function of cosmic variance measured from \texttt{UniverseMachine}, for two different mass cuts. A positive skew indicates a distribution with a high-mass tail, with the median lying below the mean, and a negative skew indicates a distribution with a low-mass tail, with the mean lying below the median. The distributions are obtained from sampling the most massive objects in 4'x4' fields in a given redshift slice, and the cosmic variance is calculated from the variance of number counts of galaxies in the same fields. The behavior is noisy, but the general trend is exactly as predicted in this paper.}
    \label{fig:simulation_validation_skew}
\end{figure*}

\begin{figure*}
    \centering
    % [trim={left bottom right top},clip]
    \includegraphics[trim={0cm 1cm 0cm 0.cm},clip,width = .9 \linewidth]{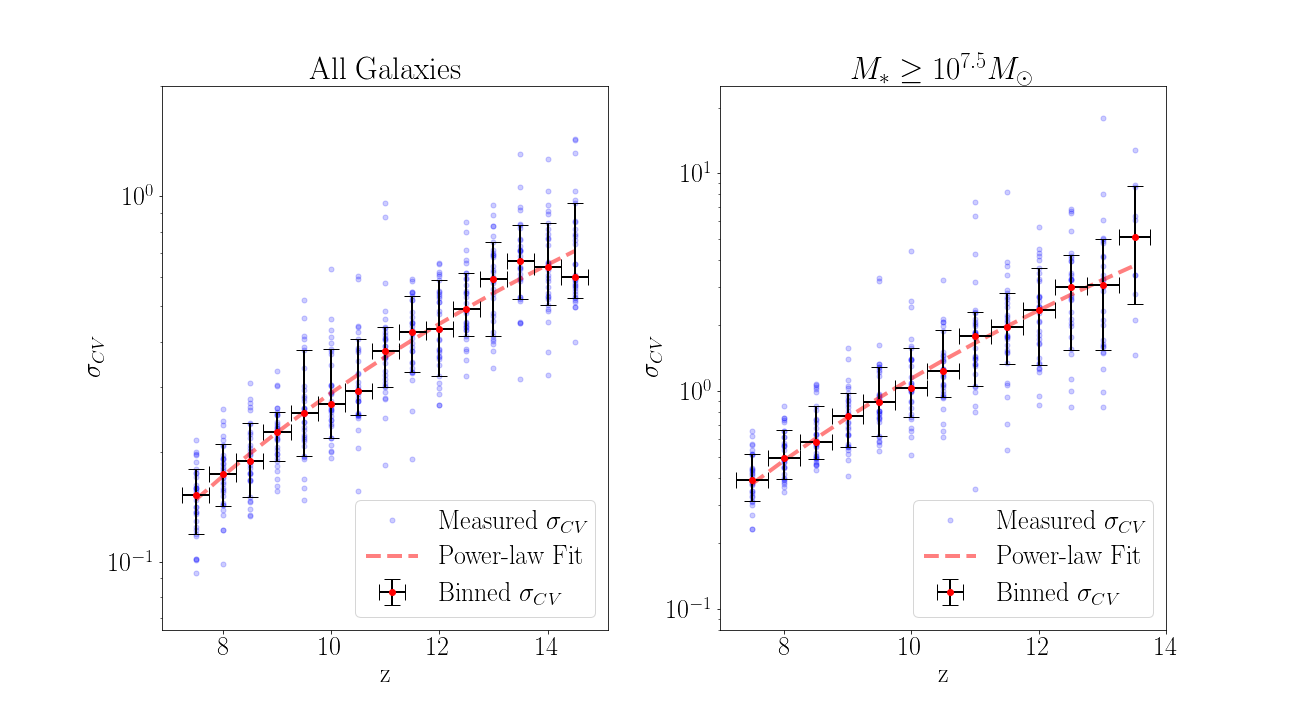}
    \caption{The measured values of $\sigma_{\mathrm{CV}}$, their binned medians and 16th and 84th percentiles from 32 \texttt{UniverseMachine} boxes, as a function of redshift. The cosmic variance is measured for two separate mass bins. However, there is very high scatter in the measured values of $\sigma_{\mathrm{CV}}$, which increases with $\sigma_{\mathrm{CV}}$ and therefore with redshift, which can also be seen in the deviation from the otherwise well-established relationship at high redshift/cosmic variance. This large scatter can be modelled from first principles, and is an inherent property of distributions with high skew that characterize galaxy number counts. Therefore, care must be taken when calibrating simulation-based cosmic variance calculators. 
    Therefore, different cosmic variance calculators calibrated to sampled fields should be expected to differ in the high cosmic variance limit.
    }
    \label{fig:cv_measurement_UM}
\end{figure*}

We test the predicted effect of the negative skew of the distribution of the mass of most massive galaxies being a consequence of high cosmic variance by sampling 4'x4' fields in redshift slices with $\Delta z = 0.5$ across the eight realizations of the four relevant \texttt{UniverseMachine} DR1 boxes from the fiducial value run \citep{Behroozi2019_UniverseMachine}. For each redshift slice, the most massive galaxy in each 4'x4' field is recorded, and the skew of the resulting distribution is measured. The numerical form of the skew is defined as in \cite{kokoska2000_StatisticalFormulae}, i.e.,

$$\gamma = \frac{\mu_3}{\sigma^3}$$

\noindent where $\mu_3 = \frac{1}{N} \Sigma (x - \langle x \rangle)^3$ is the third moment of the number count distribution.
A positive skew ($\gamma > 0$) indicates a right-skewed (heavy high-mass tail) distribution, with the median lying below the mean, and a negative skew indicates a left-skewed (heavy low-mass tail) distribution, with the mean lying below the median. Our simple model predicts that the skew goes from positive to negative as cosmic variance increases. As can be seen in Figure \ref{fig:simulation_validation_skew}, this is indeed observed in \texttt{UniverseMachine}, showing that the effects of clustering are correctly captured by the simple model. The effect is robust to mass cuts. Although there is a strong average relation between skew and cosmic variance, the behavior is by no means deterministic, as seen by the many points which diverge significantly from this mean relationship. This is due to the high effective variance associated with estimating the cosmic variance from sampling fields in a cosmological simulation with a non-Gaussian underlying mass distribution, which is discussed in the following section.

\subsection{The Variance of the Cosmic Variance}
\label{subsec:variance_of_the_variance}

Different cosmic variance calculators can yield very different results (see Figures \ref{fig:poisson_cv} - \ref{fig:bluetides_noise_poisson_cv}), especially in the high cosmic variance limit. One possible reason for these differences is likely to be the different mappings between galaxy and halo properties assumed in calibrating the different calculators. This introduces a \textit{systematic} variance in the cosmic variance.

However, here we derive and measure another, so far unrecognized, source of the differences, which is the spread/variance in the estimated cosmic variance. Therefore, this is \textit{statistical} variance in the measured cosmic variance, which arises from the properties of number count distributions that galaxies follow. As a demonstration, the estimated cosmic variance from the 32 separate \texttt{UniverseMachine} boxes used in this paper is shown in Figure \ref{fig:cv_measurement_UM}, for two separate mass limits. 

The statistical spread in the cosmic variance is visibly quite large, on the order of 1 dex for the high cosmic variance limit. This can be formalized by estimating the variance on the sample\footnote{A quantity estimated from sampling is indicated by a $\hat{~}$, e.g. an estimate of $x$ is written as $\hat{x}$. } cosmic variance, i.e.,

$$
 V\left(\widehat{\sigma_{\mathrm{CV}}^2}\right)
$$

\noindent which in turn can be used to get the more useful quantity,

$$
\sigma_{\widehat{\sigma_{\mathrm{CV}}}} = \frac{\sqrt{ V\left(\widehat{\sigma_{\mathrm{CV}}^2}\right)}}{2 \widehat{\sigma_{\mathrm{CV}}}}
$$

\noindent Making this estimate is significantly complicated by the fact that $\sigma_{\mathrm{CV}}$ is a function of both the variance and mean of the galaxy number counts, but the derivation can be done using error propagation \citep{barlow1993_statistics}. The full derivation is presented in Appendix \ref{sec:appendix_derivation}. For a number count distribution with significant positive skew, the scaling of the variance of the cosmic variance with cosmic variance is found to be

\begin{equation}
    \label{eq:sigma_on_sigma}
    \sigma_{\widehat{\sigma_{\mathrm{CV}}}} \propto \widehat{\sigma_{\mathrm{CV}}} \sqrt{\widehat{\sigma_{\mathrm{CV}}}^2 + 1}
\end{equation}

\begin{figure}
    \centering
    % [trim={left bottom right top},clip]
    \includegraphics[trim={0cm 0cm 0cm 0.cm},clip,width = \linewidth]{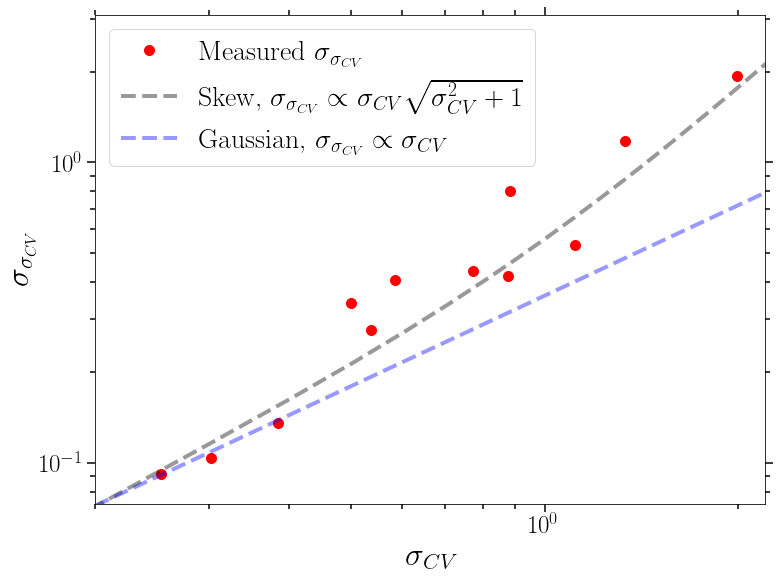}
    \caption{The variance of the measured cosmic variance ($\sigma_{\widehat{\sigma_{\mathrm{CV}}}}$) as a function of the mean measured cosmic variance ($\widehat{\sigma_{\mathrm{CV}}}$) at a given redshift, from sampled 4'x4' fields in \texttt{UniverseMachine}. The means and variances are taken across the 32 separate \texttt{UniverseMachine} boxes. The scaling of $\sigma_{\widehat{\sigma_{\mathrm{CV}}}}$ with $\widehat{\sigma_{\mathrm{CV}}}$ clearly follows that of a skewed distribution, highlighting the importance of taking into account the skew. Two similar figures for different field sizes are shown in Appendix \ref{sec:appendix_derivation}.}
    \label{fig:sigma_sigma_cv}
\end{figure}

\noindent The fit of Equation \ref{eq:sigma_on_sigma} to the observed variance is a significant improvement over what the result would be if the underlying true distribution of galaxy number counts was Poissonian or Gaussian, as can be seen in Figure \ref{fig:sigma_sigma_cv}. Notably, the fractional error is heteroscedastic. Satisfyingly, Equation \ref{eq:sigma_on_sigma} reduces to the expected $\sigma_{\widehat{\sigma_{\mathrm{CV}}}}$ for a Gaussian, 

\begin{equation}
    \label{eq:gaussian_limit}
    \sigma_{\widehat{\sigma_{\mathrm{CV}}}} \propto \widehat{\sigma_{\mathrm{CV}}}
\end{equation}

\noindent in the limit that $\widehat{\sigma_{\mathrm{CV}}} \rightarrow 0$ and $\hat{\mu} \rightarrow \infty$, which is exactly when the Gaussian assumption would be proper. It should be noted that the large statistical spread on the cosmic variance is directly connected to the underlying number count distribution having a large positive skew, as is shown in Appendix \ref{sec:appendix_derivation}.

Since this large statistical dispersion has not been taken into account when fitting previous cosmic variance calculators, it should perhaps not be surprising that they differ enormously in the high cosmic variance limit, although this may also be explained through other differences, such as differences in clustering, halo mass-stellar mass relation, or cosmology, used in different theoretical models \citep[see e.g. \S 4 of][for a full discussion]{Bhowmick2020}.

This also affects the predicted relation between cosmic variance and any other quantity. Even if the true relation is very tight, the relation will seem noisy if the estimate of the cosmic variance is very noisy. This is one of the origins of the seemingly high scatter in the $\sigma_{\mathrm{CV}}$-skew relation shown in Figure ~\ref{fig:simulation_validation_skew}. Any single estimate of cosmic variance, especially if it is high, should therefore not be given a high weight.
This behavior is somewhat intuitive since estimating the variance of a distribution with high variance, and therefore a large spread in values, should be more difficult than estimating the variance of a narrower distribution. This is in large part due to outliers being common in the high variance limit, and since the variance is dominated by outliers, the variance of high-variance distributions is hard to estimate. A possible way to minimize this issue is to use more robust estimators, like the median and interquartile ranges, for fitting cosmic variance relationships in the future. However, the increased robustness of these estimators does not fully mitigate the issue, as can be seen in Figure \ref{fig:cv_measurement_UM}. 

\subsection{The Distribution of Most Massive Galaxies in {\normalfont 
 \texttt{UniverseMachine}}}

 \begin{figure*}
    \centering
    % [trim={left bottom right top},clip]
    \includegraphics[trim={0cm 0cm 0cm 0.cm},clip,width = 1.\linewidth]{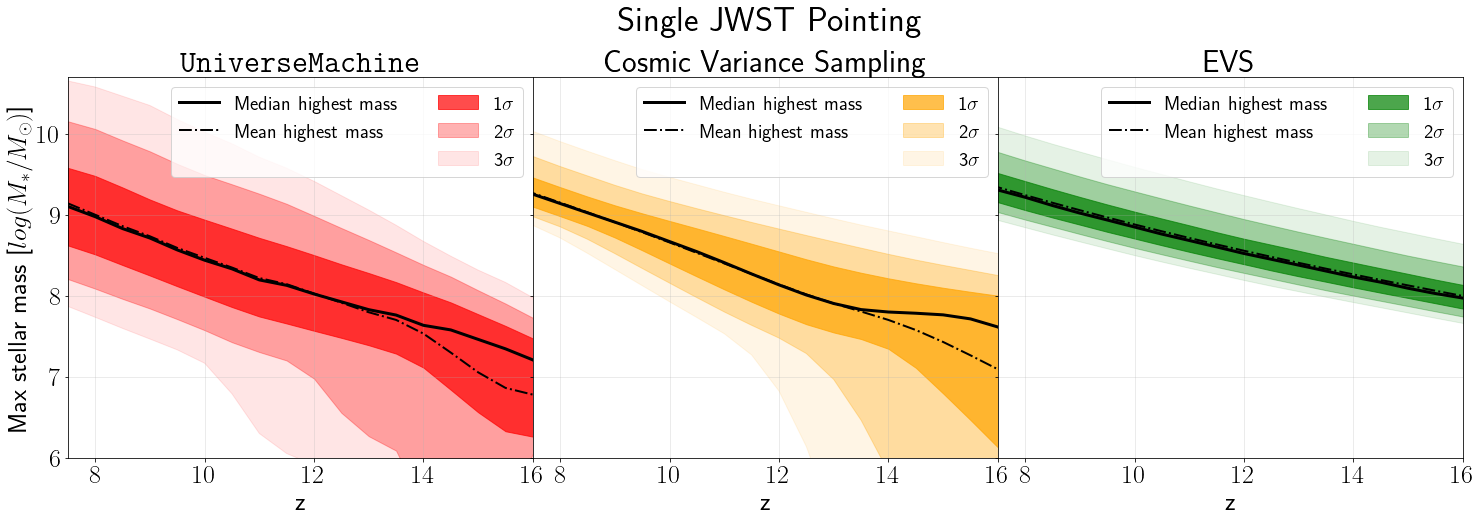}
    \caption{The probability distribution of the most massive galaxy in a single \textit{JWST} pointing having mass $M_*$, as a function of redshift. Distributions are calculated from \texttt{UniverseMachine}, for our simple cosmic variance sampling model, and EVS. The change in the shape of the \texttt{UniverseMachine} - distribution with redshift is noticeably different from that predicted by EVS, and more similar to the predictions of the simple cosmic variance model, with a strong negative skew at $z\geq11$.}
    \label{fig:um_demonstration}
\end{figure*}

Since the validity of some of the assumptions made in the simple model applied here has now been shown, a natural next point is to investigate if the same effects are observed in \texttt{UniverseMachine}. This can be tested by stacking samples of the most massive galaxy in partitioned fields in each \texttt{UniverseMachine} simulation box to obtain the distribution of the mass of the most massive galaxies. The results of doing so can be seen in Figure \ref{fig:um_demonstration}. We clearly observe the predicted effect from the simple model presented in this paper, i.e. the change in the skew of the distribution of the mass of the most massive galaxy from a high-mass skew at lower redshift/cosmic variance to a low-mass skew at high redshift/cosmic variance. \texttt{UniverseMachine} clearly shows this change, with the onset of the heavy, low-mass tail starting to happen at $z \approx 11$, which is when the median most massive galaxy starts lying above the mean most massive galaxy, as can be seen in Figure \ref{fig:um_demonstration}. The two then start diverging at $z \gtrsim 13$, with the high-mass contours becoming more compressed, and the low-mass tail becoming longer with increasing redshift, exactly as predicted by our simple model.

However, at low redshifts/cosmic variance, both EVS and our simple model underestimate the width of the distribution. This is most likely due to each approach ignoring different contributions to the total variance, which will be discussed below.

\section{Discussion}
\label{sec:discussion}

Our simple model provides a good estimate of the extent to which cosmic variance impacts the distribution of the single most massive galaxy observed in a given field. However, the exact distributions, as discussed by \cite{Lovell2023_EVS}, should only be considered with the many uncertainties that impact the model in mind. In particular, the stellar baryon fraction is a critical and nearly unconstrained parameter (modulo the upper limit arising from the limited supply of baryons in the Universe). The degree of inferred tension with a particular model is therefore highly dependent on the assumed $\epsilon_*$ \citep{Steinhardt2016_impossiblyEarly, Dekel2023_bursts}.

However, these simple models are still highly useful in investigating the \textit{relative} impact of different parts of the underlying models. In this work, it has been shown that the impact of cosmic variance moves the mean of the distribution of the mass of the most massive galaxy in a given field on the order of more than a dex for regions dominated by field-to-field variance, as well as dramatically changing the shape of the distribution. These effects then change the magnitude of the tension between observed galaxy masses and our theoretical models by up to $1 \sigma$. However, as can be seen in Figures \ref{fig:poisson_cv} - \ref{fig:bluetides_noise_poisson_cv}, the exact redshifts where these effects become significant, as well as their magnitude is highly dependent on the exact method adopted to calculate the cosmic variance. This highlights the importance of a better understanding of cosmic variance in the high redshift limit, where all approaches currently extrapolate beyond the support of the redshift intervals on which they were validated. As shown in this paper, it is also important to incorporate the statistical ``variance on the variance'' resulting from the skewed distribution of galaxy number counts (see \S \ref{subsec:variance_of_the_variance}).

However, no matter how one considers the cosmic variance, it leads to increased tension, and is not a solution to the problem of over-luminous early galaxies as has been suggested \citep{McCaffrey23_noTension}.

We must now consider the origin of the spread in the distribution of the mass of the most massive galaxies, $\sigma_M$ (note that this $\sigma_M$ is not the spread in the number count distribution), and which other possible sources of variance can reconcile the differences in $\sigma_M$ from the different approaches. There is one principal source of variance not included in our base model which can explain these differences.

\begin{figure*}
    \centering
    % [trim={left bottom right top},clip]
    \includegraphics[trim={0cm 0cm 0cm 0.cm},clip,width = 1.\linewidth]{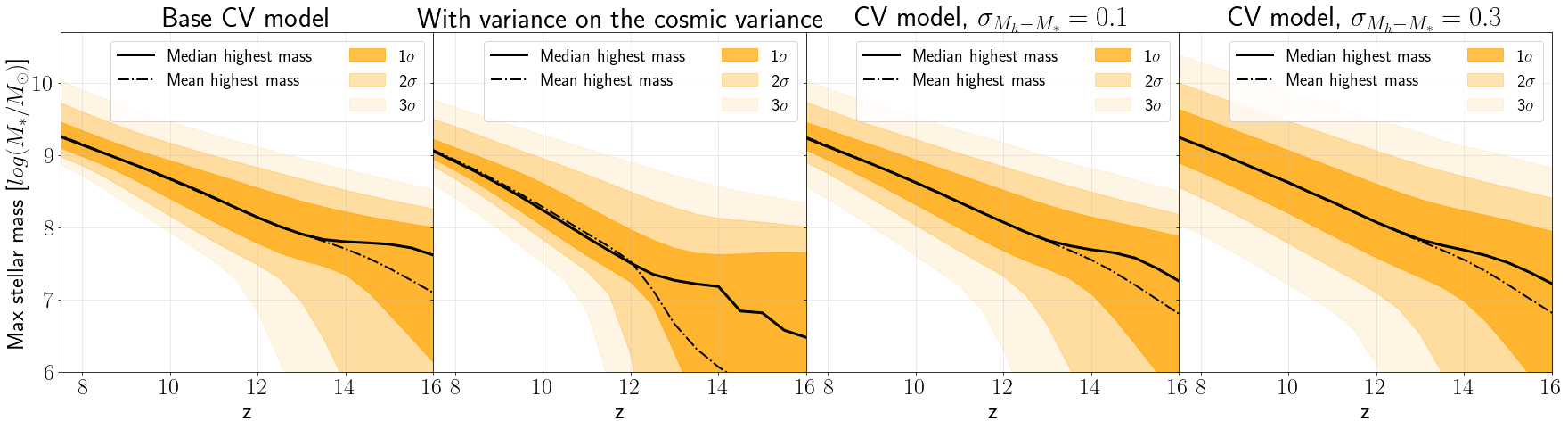}
    \caption{The probability distribution of the most massive galaxy in a single \textit{JWST} pointing having mass $M_*$, as a function of redshift, similar to Figure \ref{fig:um_demonstration}, but with significant modifications to showcase how different uncertain aspects of the modelling will affect the results. The first panel shows the results for $\sigma_{\mathrm{CV}}$ varied with the uncertainty given in Eq. \ref{eq:sigma_on_sigma} and shown in Figure \ref{fig:sigma_sigma_cv}. The introduction of this additional variance clearly makes the effect even stronger. The rightmost two panels show the effect of adding a scatter of $\sigma_M$ of 0.1 or 0.3 dex to the halo mass - stellar mass relation. The scatter here mainly effectively adds in quadrature to the already existing scatter, impacting only the $z\leq 12$ regime.}
    \label{fig:var_on_var_and_scatter}
\end{figure*}

The difference with \texttt{UniverseMachine} is mainly due to the additional scatter in the relation between halo mass and stellar mass, which was not included in our base model. However, since the model is flexible enough to include any halo-to-stellar-mass relation, we can test including it here, and thereby isolate the effect of adding scatter. This scatter is on the order of 0.1-0.3 dex \citep{Jespersen2022}, so especially at low redshift, this is a significant addition. We try including this scatter in our model, the results of which are shown in Figure \ref{fig:var_on_var_and_scatter}, and with this scatter, our results match those from \texttt{UniverseMachine} quite well. This scatter, which we will call $\sigma_{\mathrm{M,M_{\mathrm{h}}-M_*}}$, is the dominant additional source of variance.
The two additional terms which set the width of the distribution of masses of the most massive galaxies are that of cosmic variance, $\sigma_{\mathrm{M,CV}}$, and from the Poisson variance, $\sigma_{\mathrm{M,Poisson}}$. 
The authors wish to stress that all $\sigma_{\mathrm{M,x}}$ (with the $M$ subscript), refer to the spread in the distribution of the masses of the most massive galaxies, \textit{not} the spread in the number count distribution.
Summarizing all of the potential contributions, we may think of the total spread in the distribution of the masses of the most massive galaxies as

$$\sigma_{\mathrm{M, total}}^2 = \sigma_{\mathrm{M,M_{\mathrm{h}}-M_*}}^2 + \sigma_{\mathrm{M,CV}}^2+\sigma_{\mathrm{M,Poisson}}^2$$

\noindent where EVS ignores the first and second term, and the basic cosmic variance model presented here ignores the first term. As can be seen in Figure \ref{fig:var_on_var_and_scatter}, the most likely origin of the additional width in the distribution of most massive galaxies observed in \texttt{UniverseMachine} at lower redshift is the assumption of a one-to-one halo to stellar mass matching in the cosmic variance model presented here. The width of the distribution of the mass of the most massive galaxies is also impacted by how strongly mass-scale correlation is enforced, with the distribution becoming narrower as this assumption is relaxed.

Cosmic variance will also be highly relevant for lensed fields like the UNCOVER survey \citep{Bezanson2022_UNCOVER_design}. Modelling the impact of cosmic variance on the most massive galaxies for these fields is even more complicated \citep{Vujeva2023_lensing} and therefore left for future work.

\subsection{Sources of Uncertainty for Galaxy Formation Models}

The distribution of the masses of the most massive galaxy in a given field is subject to many uncertainties and biases due to the modelling assumptions. Since the approach presented in this paper estimates the magnitude of the change in the distribution of most massive galaxies introduced by considering cosmic variance, we can now estimate the relative impact of different modelling choices, and determine the regimes in which the uncertainty on each dominates.

For pencil-beam surveys at intermediate redshifts ($4<z<8$), the main source of subjectivity comes from the adopted stellar baryon fraction \citep{Finkelstein2015_SBF, Steinhardt2016_impossiblyEarly, Boylan-Kolchin23_early_galaxies, Dekel2023_bursts}, which can change the galactic stellar masses by up to 0.5 dex in either direction. As discussed above, another assumption that adds an uncertainty of between 0.1-0.3 dex \citep{Jespersen2022}, is the assumption that all galaxies can be matched to halos in a monotonic fashion using only halo mass. We have tested including this effect to investigate its relative importance, which is shown in Figure \ref{fig:var_on_var_and_scatter}. This additional scatter effectively adds in quadrature to the already existing scatter, and since the pre-existing scatter exceeds any realistic scatter one could introduce into the stellar-to-halo mass relationship at $z\geq 12$, this mainly impacts the $z\leq 12$ regime. For large, high-redshift fields ($z\geq8$), where cosmic variance is low (square degree or many independent pointings), the choice of parametrization of halo mass function becomes increasingly impactful, with differences in number densities of up to a dex \citep{Yung23_highz}. Therefore, it is increasingly important to use fitting functions appropriate in the high-z regime, or directly use the results of an N-body simulation. All of these discussed effects are also relevant to distributions of most massive galaxies derived with EVS. The cosmological baryon fraction may also not be universal \citep{Crain2007_baryon_frac}, which would globally lower the stellar masses. However, this would also mean that lower mass galaxies reside in more massive halos than what has been assumed here, which would make them more strongly affected by cosmic variance.

For surveys of similar size to or smaller than CEERS/JADES, the main source of uncertainty at ultra-high redshifts ($z \geq 10$) comes from the cosmic variance. The distributions of most massive galaxies calculated with different calculators can vary by extreme amounts at fixed redshift. For example, for a CEERS-like field, the average most massive galaxy at $z = 15$ has a mass of $ 10^{8.1} M_{\odot}$ if the cosmic variance of \texttt{UniverseMachine} is adopted, but a mass of $ 10^{7.1} M_{\odot}$ if using the \texttt{BlueTides} calculator of \cite{Bhowmick2020}. This difference can be thought of as a \textit{systematic} variance on the cosmic variance. For a single deep pointing, these differences are obviously even larger. The differences between calculators arise in part due to the newly introduced statistical ``variance on the variance'', which has so far been neglected, and from the uncertainty in linking observable galaxy properties (such as luminosity or, more indirectly, stellar mass) with halo mass, which translates to an uncertainty on the bias (or clustering strength) of a given observed population. We can also investigate the impact of only the statistical variance on the cosmic variance alone, simply by sampling different cosmic variances from the samples from \texttt{UniverseMachine}. The ``variance on the variance'' is incorporated by sampling five different $\sigma_{\mathrm{CV}}$ at the $\{0.17, 0.33, 0.5, 0.67, 0.83\}$ percentiles\footnote{Five equally sampled steps in the ]0,1[ interval.} of the distributions of possible $\sigma_{\mathrm{CV}}$'s. In the 5-step summary in \S \ref{subsec:max_val_dist_procedure}, this corresponds to repeating steps 3 - 5 with five different $\sigma_{\mathrm{CV}}$ and combining the final results. The results of this procedure can be found in Figure \ref{fig:var_on_var_and_scatter}, which shows the probability contours being even wider when taking the indeterminable nature of the cosmic variance into account.

The impact of the uncertainty on the cosmic variance itself is likely even larger when the uncertainties discussed above are propagated. This means that for the deepest \textit{JWST} surveys, which push the boundary of the current redshift limit, cosmic variance will be the biggest source of uncertainty in determining if a given massive galaxy is in tension with theoretical models. We therefore need more robust estimates of cosmic variance at high redshift. Upcoming JWST surveys will be able to empirically provide a calibration basis for new cosmic variance calculators in the $z \geq 10$ regime, although these calibrations should be attentive to the large inherent uncertainties on cosmic variance estimates in the high cosmic variance regime, as shown in \S \ref{subsec:variance_of_the_variance} and Appendix \ref{sec:appendix_derivation}.

Naturally, many other theoretical and observational uncertainties also need to be reduced to robustly constrain galaxy models, especially the photometric low-redshift interlopers \citep{ArrabalHaro2023_spec_confirmation_jwst} and the uncertainties in determining stellar masses \citep{Steinhardt2023_IMF_template}. However, as can be seen in Figure \ref{fig:singleobject}, even spectroscopically confirmed objects still imply tension with galaxy formation models. It should be noted that the full extent of the uncertainty associated with determining stellar mass at high redshift is still not fully known at this point. Determining solutions to these issues will require hard, but exciting, endeavours, but there is scarcely any passion without struggle \citep{camus1942_sisyphus}. 

\section{Conclusion}
\label{sec:conclusion}

In this work, the impact of cosmic variance on the probability of finding a single extremely massive galaxy in a given survey has been investigated. At $z = 14$, when comparing the cosmic variance sampling model prediction to the EVS prediction, the expected mass of the mean most massive galaxy is lower by almost 1 dex for a single \textit{JWST} pointing and 0.5 dex lower for a CEERS/JADES-like survey. The median is slightly more robust, changing less due to the impact of cosmic variance. Including cosmic variance increases the tension between galaxy formation models and the masses of observed CEERS/JADES galaxies by $\approx 1 \sigma$. These differences grow significantly if we assume the cosmic variance from \texttt{BlueTides} instead of the cosmic variance from \texttt{UniverseMachine}, highlighting the importance of improved cosmic variance calculators. 

When cosmic variance is larger than Poisson variance, we find that the shape of the distribution of most massive galaxies changes from a positively skewed (high-mass tail) distribution to a negatively skewed (low-mass tail) distribution, which results in the mass of the median most massive galaxy being higher than the average most massive galaxy. When the total variance is larger than the average number of galaxies in a field, the mass distribution becomes heavy-tailed towards the low-mass end, and the mean and median diverge strongly. These effects make cosmic variance the largest source of uncertainty for deep, ultra-high redshift fields. For lower redshifts ($z\leq12$), scatter in the relation between halo and stellar mass is the dominant source of uncertainty.

We have furthermore shown that cosmic variance estimates have significantly higher statistical dispersion than previously thought. This ``variance on the variance'' is a direct result of the highly skewed nature of the galaxy number count distribution. The increased variance on cosmic variance estimates should be taken into account when calibrating future cosmic variance calculators, a necessary task for robustly constraining galaxy models at high redshift. This additional variance greatly impacts the expected mass of the most massive galaxy in a given field.

Our predictions and assumptions have been validated using the \texttt{UniverseMachine} simulation suite, showing good agreement.\\
Since evaluating any tension based on single rare objects is likely to be sensitive to large or underestimated errors on their modelled properties, it is generally preferable to use the properties of larger samples to constrain models. However, if one wishes to use extreme objects in this manner, our results show that it is critical to properly include the effects of cosmic variance.

\section{Acknowledgements}

The authors thank Brant Robertson, Michael Boylan-Kolchin, ChangHoon Hahn, Jiaxuan Li, Adrian Bayer, Jeff Shen, Zachary Hemler, Philippe Yao, Marielle Côté-Gendreau and Yilun Ma for helpful comments. The authors would like to thank John F. Wu, Francisco Villaescusa-Navarro, Tjitske Starkenburg and Peter Behroozi for organizing the 2023 KITP program\textit{ Building a Physical Understanding of Galaxy Evolution with Data-driven Astronomy}.
The authors would also like to thank the
anonymous reviewer for their helpful and concise comments, which greatly clarified the contents of this paper.
This research was supported in part by grant NSF PHY-1748958 to the Kavli Institute for Theoretical Physics (KITP), which made the completion of this work possible. The Cosmic Dawn Center (DAWN) is funded by the Danish National Research Foundation under grant No. 140.  

\bibliographystyle{aasjournal}
\bibliography{main,extremes}

%%%%%%%%%%%%%%%%%%%%%%%%%%%%%%%%%%%%%%%%%%%%%%%%%%

%%%%%%%%%%%%%%%%% APPENDICES %%%%%%%%%%%%%%%%%%%%%

\appendix

\section{Full Derivation of the Variance on the Cosmic Variance}
\label{sec:appendix_derivation}

Since cosmic variance is derived via sampling, there will be some variance due to the sampling. We therefore wish to derive 

\begin{equation}
\label{eq:var_on_sig_cv}
     V\left(\widehat{\sigma_{\mathrm{CV}}}\right), 
     ~ \widehat{\sigma_{\mathrm{CV}}^2}=\frac{\hat{\sigma}^2-\hat{\mu}}{\hat{\mu}^2}=\frac{\hat{\sigma}^2}{\hat{\mu}^2}-\frac{1}{\hat{\mu}}
\end{equation}

where $\hat{\mu} = \langle \hat{N} \rangle$, and $\hat{\sigma} = \langle \hat{N}^2 \rangle - \langle \hat{N} \rangle^2$, following Equation \ref{eq:cosmic_variance_defintion}. $\hat{x}$ designates the estimated $x$, so $\hat{\sigma}$ is the sample standard deviation, often also called $s$ \citep{barlow1993_statistics}, not the true standard deviation $\sigma$. 

Since $\widehat{\sigma_{\mathrm{CV}}^2}$ is a composite quantity, propagation of errors can be applied to get the total variance. For a function $f(a, b)$, with uncertain a, b,

\begin{equation}
V(f(a,b))=\left(\frac{\partial f}{\partial a}\right)^2 \sigma_a^2+\left(\frac{\partial f}{\partial b}\right)^2 \sigma_b^2+2 \frac{\partial f}{\partial a} \frac{\partial f}{\partial b} \mathrm{cov}(a, b)
\end{equation}

see \cite{barlow1993_statistics}, page 57.

This means that the variance of the sample cosmic variance can be estimated as 

\begin{equation}
    \label{eq:error_prop_sig_cv}
    V(\widehat{\sigma_{\mathrm{CV}}^2})=\left(\frac{\partial \widehat{\sigma_{\mathrm{CV}}^2}}{\partial \hat{\sigma}^2}\right)^2 V(\hat{\sigma}^2)+\left(\frac{\partial \widehat{\sigma_{\mathrm{CV}}^2}}{\partial \hat{\mu}}\right)^2 V(\hat{\mu})+2 \frac{\partial \widehat{\sigma_{\mathrm{CV}}^2}}{\partial \hat{\sigma}^2} \frac{\partial \widehat{\sigma_{\mathrm{CV}}^2}}{\partial \hat{\mu}} \mathrm{cov}(\hat{\sigma}^2, \hat{\mu})
\end{equation}

The derivatives are easily found.

\begin{equation}
    \label{eq:derivative_sig}
    \frac{\partial \widehat{\sigma_{\mathrm{CV}}^2}}{\partial \hat{\sigma}^2} = \frac{1}{\hat{\mu}^2}
\end{equation}

\begin{equation}
    \label{eq:derivative_mu}
    \frac{\partial \widehat{\sigma_{\mathrm{CV}}^2}}{\partial \hat{\mu}} = \frac{\hat{\mu}-2 \hat{\sigma}^2}{\hat{\mu}^3}
\end{equation}

The variances on $\hat{\sigma}$ and $\hat{\mu}$ are also known \citep{barlow1993_statistics, O'Reilly2014_statistics_moments}.

\begin{equation}
    \label{eq:sig_variance_theory}
  V\left(\hat{\sigma}^2\right)=\frac{\hat{\sigma}^4}{n^3}\left(\kappa \cdot(n-1)^2-(n-1)(n-3)\right)   
\end{equation}

where $n$ in is the number of sampled fields, and $\kappa$ is the \textbf{kurtosis} (the normalized fourth moment). The variance on the sample mean is

\begin{equation}
\label{eq:mu_variance}
 V(\hat{\mu})=\frac{\hat{\sigma}^2}{n}    
\end{equation}

The covariance between the sample variance and sample mean is also known
\begin{equation}
    \label{eq:covariance}
    \mathrm{cov}(\hat{\sigma}^2, \hat{\mu}) = \frac{\gamma \hat{\sigma}^3}{n}
\end{equation}

where $\gamma$ is the \textbf{skew} (the normalized third moment) of the sampled distribution.\\

Equations \ref{eq:error_prop_sig_cv} - \ref{eq:covariance} can be combined to give the total variance on the $\widehat{\sigma_{\mathrm{CV}}^2}$

\begin{equation}
V\left(\widehat{\sigma_{\mathrm{CV}}^2}\right) =\frac{\hat{\sigma}^4}{\hat{\mu}^4 n^3}\left(\kappa (n-1)^2-(n-1)(n-3)\right)+\left(\frac{2 \hat{\sigma}^2-\hat{\mu}}{\hat{\mu}^3}\right)^2 \frac{\hat{\sigma}^2}{n} -2\left(\frac{2\hat{\sigma}^2-\hat{\mu}}{\hat{\mu}^5}\right) \frac{\hat{\sigma}^3 \gamma}{n}    
\end{equation}

For the sake of demonstration, we choose the kurtosis and skew of the gamma distribution to carry forward, 

$$\kappa=\frac{6 \hat{\sigma}^2}{\hat{\mu}^2}+3 $$
$$ \gamma=\frac{2 \hat{\sigma}}{\hat{\mu}}$$

Although this is for a specific distribution, it is a general property of skewed distribution that $\kappa \propto \frac{\sigma^2}{\mu^2}$ and $\gamma \propto \frac{\sigma}{\mu}$ \citep{Pearson1916_skew}. While the exact pre-factors can therefore change with the assumed distribution, the scalings remain the same. The resulting scaling is the same if a negative binomial is assumed.\\

Inserting the above, we get,

\begin{equation}
    V\left(\widehat{\sigma_{\mathrm{CV}}^2}\right)=\frac{\hat{\sigma}^4}{\hat{\mu}^4} \left( \frac{1}{n^3} \left((n-1)^2\left(\frac{6 \hat{\sigma}^2}{\mu^2}+3\right)-(n-1)(n-3)\right) +\left(\frac{2 \hat{\sigma}^2-\hat{\mu}^2}{\hat{\mu}^2}\right)^2 \frac{\hat{\mu}^2}{\hat{\sigma}^2 n}-\frac{4}{n}\left(\frac{2 \hat{\sigma}^2-\hat{\mu}}{\mu^2}\right)\right)    
\end{equation}

If we optimistically assume that $n \rightarrow \infty$, i.e. that we have access to a large number of samples (this is actually a good approximation for $n \gtrsim 10$), we can condense this even further.

\begin{equation}
    \label{eq:before_sigcv_transfor}
    V\left(\widehat{\sigma_{\mathrm{CV}}^2}\right)=\frac{\hat{\sigma}^4}{\hat{\mu}^4 n} \left( \frac{6 \hat{\sigma}^2}{\hat{\mu}^2}+2 +\left(\frac{2 \hat{\sigma}^2-\hat{\mu}^2}{\hat{\mu}^2}\right)^2 \frac{\hat{\mu}^2}{\hat{\sigma}^2}-4\left(\frac{2 \hat{\sigma}^2-\hat{\mu}}{\hat{\mu}^2}\right)\right)    
\end{equation}

To gain further insight, we can cast Equation \ref{eq:before_sigcv_transfor} in terms of $\sigma_{\mathrm{CV}}$ by noting that,

$$
\widehat{\sigma_{\mathrm{CV}}^2}=\frac{\hat{\sigma}^2-\hat{\mu}}{\hat{\mu}^2}=\frac{\hat{\sigma}^2}{\hat{\mu}^2}-\frac{1}{\hat{\mu}},
$$
$$ \frac{\hat{\sigma}^2}{\hat{\mu}^2}=\widehat{\sigma_{\mathrm{CV}}^2}+\frac{1}{\hat{\mu}}$$

and,
$$
\frac{2 \hat{\sigma}^2-\hat{\mu}}{\hat{\mu}^2}=\frac{2 \hat{\sigma}^2}{\hat{\mu}^2}-\frac{1}{\hat{\mu}}=2 \widehat{\sigma_{\mathrm{CV}}^2}+\frac{1}{\hat{\mu}} 
$$

Inserting these in Equation \ref{eq:before_sigcv_transfor}, we get,

\begin{equation}
V\left(\widehat{\sigma_{\mathrm{CV}}^2}\right)=\frac{1}{n}\left(\widehat{\sigma_{\mathrm{CV}}^2}+\frac{1}{\hat{\mu}}\right)^2\left(6\left(\widehat{\sigma_{\mathrm{CV}}^2}+\frac{1}{\hat{\mu}}\right)+2+\frac{\left(2 \widehat{\sigma_{\mathrm{CV}}^2}+\frac{1}{\hat{\mu}}\right)^2}{\left(\widehat{\sigma_{\mathrm{CV}}^2}+\frac{1}{\hat{\mu}}\right)} -8\widehat{\sigma_{\mathrm{CV}}^2}-\frac{4}{\hat{\mu}} \right)   
\end{equation}

Assuming, again optimistically, that we have a well-measured field with many galaxies, i.e.  $\frac{1}{\hat{\mu}} \rightarrow 0$, the equation reduces to

\begin{equation}
    \label{eq:almost_there}
    V\left(\widehat{\sigma_{\mathrm{CV}}^2}\right) = \frac{2 \widehat{\sigma_{\mathrm{CV}}}^4}{n} (\widehat{\sigma_{\mathrm{CV}}^2}+1)
\end{equation}

Using Equation 4.9 from \cite{barlow1993_statistics},

\begin{equation}
    V\left(\widehat{\sigma_{\mathrm{CV}}}\right) = \frac{V\left(\widehat{\sigma_{\mathrm{CV}}^2}\right)}{4 \widehat{\sigma_{\mathrm{CV}}^2}} = \frac{\widehat{\sigma_{\mathrm{CV}}^2}}{2n} (\widehat{\sigma_{\mathrm{CV}}^2}+1)
\end{equation}

We then get the expected standard deviation on $\widehat{\sigma_{\mathrm{CV}}}$ as,
\begin{equation}
\label{eq:final_uncertainty}
\sigma_{\widehat{\sigma_{\mathrm{CV}}}} = \sqrt{V\left(\widehat{\sigma_{\mathrm{CV}}}\right)} = \frac{1}{\sqrt{2n}} \widehat{\sigma_{\mathrm{CV}}} \sqrt{\widehat{\sigma_{\mathrm{CV}}^2}+1}
\end{equation}

For $\widehat{\sigma_{\mathrm{CV}}} \gg 1$, the sample variance on the cosmic variance is therefore expected to be extremely large, even when having access to a very large amount of samples, since the respective scalings are very different. $\widehat{\sigma_{\mathrm{CV}}} \gg 1$ is the limit in which the underlying distribution of galaxy number counts is highly skewed, so the high uncertainty on the cosmic variance is a result of the high skew of the number count distribution. This makes intuitive sense, since estimating the variance of a distribution with high skew and high variance, is dominated by outlier draws, making the estimated variance highly uncertain.

Notably, in the $\widehat{\sigma_{\mathrm{CV}}} \ll 1$ limit, Equation \ref{eq:final_uncertainty} reduces to the relation that $\sigma_{\widehat{\sigma_{\mathrm{CV}}}}$ would have if the underlying distribution of galaxy number counts were Gaussian, 

$$\sigma_{\widehat{\sigma_{\mathrm{CV}}}} = \frac{\widehat{\sigma_{\mathrm{CV}}}}{\sqrt{2n}}$$

Which is known from \cite{barlow1993_statistics}.

\begin{figure}[h]
    \centering
    % [trim={left bottom right top},clip]
    \includegraphics[trim={0cm 0cm 0cm 0.cm},clip,width = 0.5\linewidth]{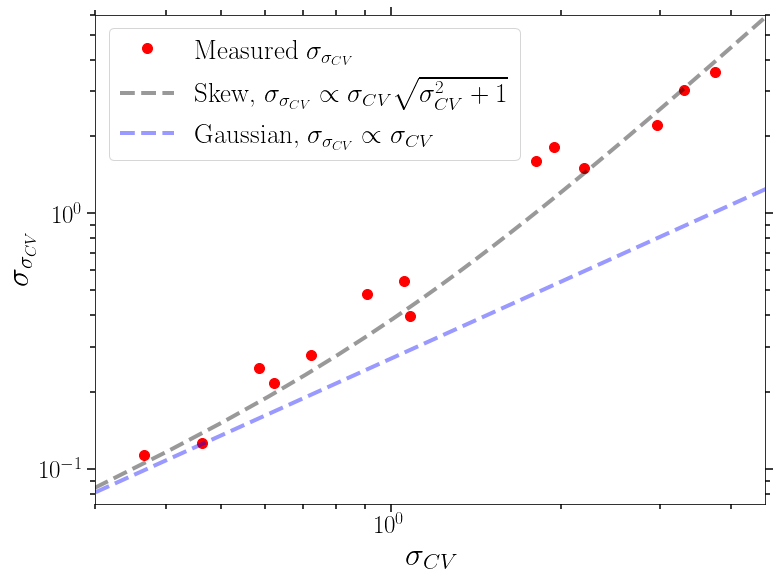}
    \caption{Same as Figure \ref{fig:sigma_sigma_cv}, but for 2'x4' fields, half the area of the fields sampled for the figure in the main text. The predicted scaling (Equation \ref{eq:final_uncertainty}) reproduces the observed variance on estimated cosmic variances surprisingly well given the many simplified assumptions made in the derivation. This effect will be impactful when calibrating a cosmic variance calculator, and its omission is likely to explain some of the differences between different cosmic variance calculators.}
    \label{fig:small_sigma_sigma_cv}
\end{figure}

\begin{figure}[h]
    \centering
    % [trim={left bottom right top},clip]
    \includegraphics[trim={0cm 0cm 0cm 0.cm},clip,width = 0.5\linewidth]{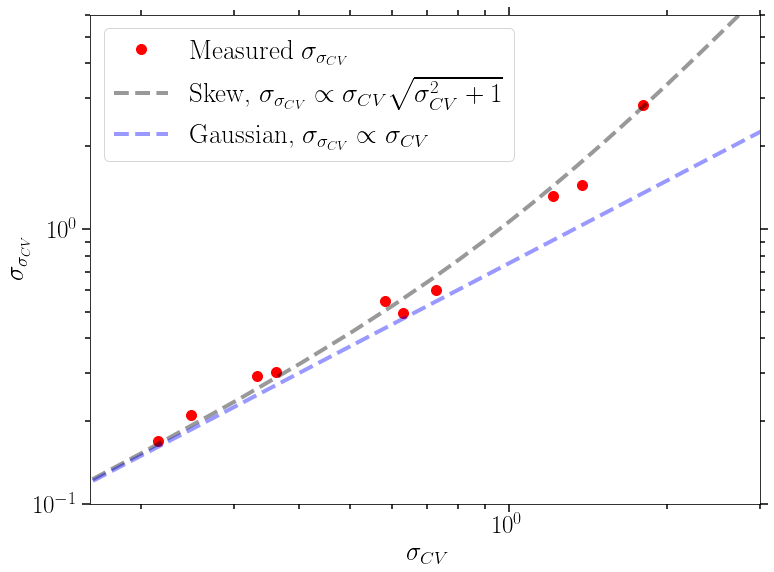}
    \caption{Same as Figure \ref{fig:sigma_sigma_cv}, but for 5'x7' fields, similar to the area of JADES/CEERS. The predicted scaling (Equation \ref{eq:final_uncertainty}) again reproduces the observed variance on estimated cosmic variances well, especially in the $\sigma_{\mathrm{CV}}>1$ limit, where the new corrections made are most significant.}
    \label{fig:JADES_sigma_sigma_cv}
\end{figure}

\section{Mass Scale Covariance}
\label{appsec:masscales}

\begin{figure}[h]
    \centering
    % [trim={left bottom right top},clip]
    \includegraphics[trim={0cm 0cm 0cm 0.cm},clip,width = 0.7\linewidth]{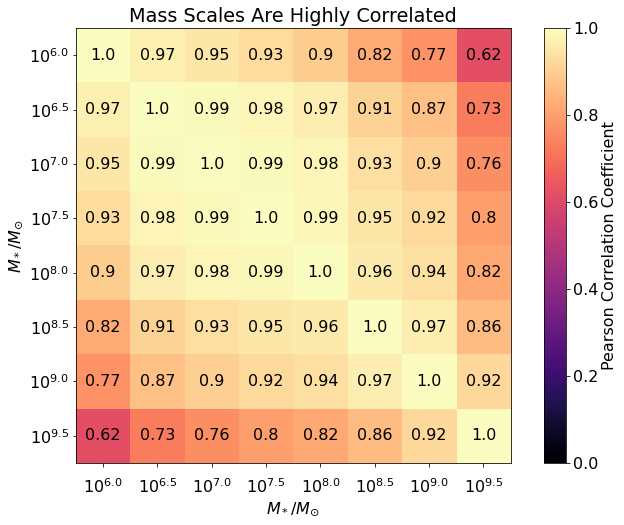}
    \caption{Linear correlation coefficients between the number counts in different mass bins sampled from subfields of the \texttt{UniverseMachine} simulations. The correlation coefficients are generally quite high, especially between neighboring bins.}
    \label{fig:mass_scale}
\end{figure}

Since our work hinges partially on the assumptions that number count fluctuations in different mass bins correlate, we validate this assumption using \texttt{UniverseMachine}, although other work has already shown that local environments impact galaxies strongly \citep{WuJespersen2023_environment, WuJespersen2024_environment}. For our test, we sample small (4'x4') fields across a subset of the boxes, across many redshifts, and compute the linear (Pearson) correlation coefficient. The results are shown in Figure \ref{fig:mass_scale}. It is clear, as is expected, that neighboring mass bins are more strongly correlated than more distant mass bins, but even across 3.5 orders of magnitude in mass, the masses remain highly covariant. This is consistent with the work of the FLARES team \citep{Thomas2023_FLARES_overdensity}, as discussed in the main text. This shows that while our assumption of fully correlated mass bins is not completely correct, it is reasonably accurate.

\end{document}